\documentclass[a4,notitlepage,12pt]{jedm}
\usepackage[table]{xcolor}
\usepackage{url}
\usepackage{hyperref}
\usepackage{amsmath}
\usepackage{commath}
\usepackage{graphicx}
\usepackage{multirow}

\usepackage{caption}
\usepackage{subcaption}
\usepackage{xspace}
\usepackage{paralist}

\newcommand{\name}{RiPLE\xspace}
\begin{document}
\title{\name: Recommendation in Peer-Learning Environments Based on Knowledge Gaps and Interests }

\date{}

\author{{\large Hassan Khosravi}\\University of Queensland\\h.khosravi@uq.edu.au \and {\large Kendra Cooper}\\Independent Scholar\\kendra.m.cooper@gmail.com \and {\large Kirsty Kitto}\\ University of Technology Sydney \\kirsty.kitto@uts.edu.au}

\maketitle
%
\begin{abstract}
Various forms of Peer-Learning Environments are increasingly being used in post-secondary education, often to help build repositories of student generated learning objects. However, large classes can result in an extensive repository, which can make it more challenging for students to search for suitable objects that both reflect their interests and address their knowledge gaps. Recommender Systems for Technology Enhanced Learning (RecSysTEL) offer a potential solution to this problem by providing sophisticated filtering techniques to help students to find the resources that they need in a timely manner. Here, a new RecSysTEL for Recommendation in Peer-Learning Environments (\name) is presented. The approach uses a collaborative filtering algorithm based upon matrix factorization to create personalized recommendations for individual students that address their interests and their current knowledge gaps. The approach is validated using both synthetic and real data sets. The results are promising, indicating \name is able to provide sensible  personalized recommendations for both regular and cold-start users under reasonable assumptions about parameters and user behavior. 
\end{abstract}

\section{Introduction}

The importance of peer-learning in post-secondary education is being increasingly recognized  \cite{Boud2014}. This has led to the creation of a number of Peer-Learning Environments that are claimed to perform many different roles. They are often designed to: engage and satisfy students by instilling ownership; help build communities and recognize participation; and can provide rich, timely peer-generated feedback \cite{BETTS2013,Coetzee2015,Wright2015}.

For example, PeerWise \cite{Denny2008} is a free web-based system in which students can create multiple-choice questions as well as answer, rate, and discuss questions created by their peers. Empowering students with environments like these offers significant benefits, as they enhance student involvement in cognitively demanding tasks rather than the more passive answering of questions. Thus, students using PeerWise are required to identify missing knowledge, diagnose misconceptions, and provide feedback to their peers in their own words. Such tasks all ultimately enhance student learning \cite{Chin2002,Rosenshine1996,Hardy2014}. 
However, as the class size of a PeerWise instance grows, so does the number of available questions on the platform. This makes it more challenging for students to select questions that best suit their current learning needs. Will students select questions that fill their current knowledge gap? Or will they just select those easy questions that make them feel like they have mastered the material that they should be learning? To thrive in learning environments of this form, students need to be able to identify the characteristics of questions that will be both interesting and the most beneficial for their current knowledge needs. But students often lack the requisite skills for making good decisions about what and how to study  \cite{beswick.rothblum.ea:psychological,biggs:what}, which can leave them undirected and wasting time. 

Recommender systems (RecSys) \cite{Ricci2011} offer a potential solution to problems of overwhelming choice by providing sophisticated filtering techniques to help people find the resources that they need. 
Specifically, as the field of recommender systems for technology enhanced learning (RecSysTEL) \cite{Manouselis2011} evolves it becomes possible to analyze the digital traces left by learners in these environments and use them to provide recommendations about resources that will most meet their learning needs and interests. 

Here, a novel RecSysTEL solution is presented that helps students to navigate in complex Peer-Learning Environments, specifically those that deal with question answering in the form provided by PeerWise. An examination in Section~\ref{sec:related work} of related work in profiling student knowledge and RecSysTEL demonstrates that there is reason to believe that progress can be made by combining more sophisticated learner profiles with RecSysTEL solutions. In Section~\ref{sec:rippleTechniques} the techniques and technologies used in the solution, \name, are introduced in addition to the problem statement. The solution itself is presented in Section~\ref{sec:riple}, and a simple example demonstrating how \name works is in Section~\ref{sec:example} Experimental validation and results are reported in Sections~\ref{sec:validation} and~\ref{sec:exploration}, followed by conclusions and a discussion of future work in Section~\ref{sec:conclusion} 

\section{Related Work}\label{sec:related work}

\subsection{Recommendation Systems for Technology Enhanced Learning}
\label{sec:RecSysTEL}

RecSysTEL is an active and rapidly evolving research field. For example, \citeN{Drachsler2015} perform an extensive classification of 82 different RecSysTEL environments, and \citeN{Erdt2015} review the various evaluation strategies that have been applied in the field. Together these articles provide comprehensive recent surveys that consider more than 200 articles spanning over 15 years. 
Here, the focus is on collaborative filtering (CF), which identifies similar users and provides recommendations based upon their usage patterns. CF has been extensively employed in RecSysTEL, and an early LAK paper, \citeN{Verbert2011} evaluated and compared the performance of different CF techniques on educational data sets, showing that the best choice of algorithm is data dependent. In a more recent study, \protect\citeN{Kopeinik2017}
also concluded that the performance of the algorithms strongly depends on the properties and characteristics of the particular dataset.
In combining educational data sets with social networks, \citeN{Cechinel2013} used CF to predict the utility of items for users based on their interest and the interest of the network of the users around them, and \citeN{Fazeli2014} proposed a graph-based approach that uses graph-walking for improving performance on educational data sets.

One important way in which RecSysTEL has been used in an educational setting is to recommend personalized learning objects. Thus, \citeN{Lemire2005} used inference rules to provide context aware recommendation on learning objects, and \citeN{Mangina2008} recommend documents and resources within e-learning environments to expand or reinforce knowledge. Interestingly, \citeN{Gomez-Albarran2009} combined content based filtering with collaborative filtering to make recommendations in a student authored repository. 
When recommending learning objects (e.g. questions) related to student knowledge gaps, \citeN{Cazella2010} provided a semi-automated, hybrid solution based on CF (nearest neighbor) and rule based filtering, while \citeN{Thai-Nghe2011} used students' performance predictions to recommend more appropriate exercises. CF 

techniques (basic, biased, and tensor matrix factorization) were used to address a number of different student behaviors and to model the temporal effect of students improving over time. Recently, \citeN{Imran2016} provided an automated solution to personalize a learning management systems (LMS) using advanced learners' profiles to encapsulate their expertise level, prior knowledge, and performance in the course. The approach used association rule mining to create the learning object recommendations. 

Matrix factorization (MF) is one of the most established techniques used in CF; however, despite its success in RecSys, MF has rarely been used in RecSysTEL. Out of the 124 papers referenced by \cite{Drachsler2015}, only two papers directly use it \cite{Salehi2013,Thai-Nghe2011}. In Section~\ref{sec:q-matrix} the discussion reveals that this is somewhat surprising; MF has been put to good use in EDM for the purpose of generating latent profiles of student expertise and so ought to combine with RecSysTEL in a straightforward manner. Indeed, the intelligent tutoring systems that appear to be utilized in that body of work could be seen as closely related to RecSysTEL, although they tend to give students less autonomy to accept or reject the pathways chosen for them \cite{chen2008}. 

MF is particularly powerful in modeling students' performance and knowledge, because it implicitly incorporates guess and slip factors as latent factors \cite{Thai-Nghe2011}. In a question answering scenario slipping refers to the situation where a student has the required skill for answering a question but mistakenly provides the wrong answer; guessing refers to the situation where a student provides the right answer despite not having the required skill for solving the problem. This is a complex task that has received significant attention in the EDM community \cite{Beck2007,Baker2008,Pardos2010}. 

\subsection{Learner Profiling Using the Q-matrix}
\label{sec:q-matrix}

In contrast to the RecSysTEL literature an ongoing program of research, spanning more than 30 years, has sought to build models of student competencies and underlying knowledge, mapping them to educational tasks. An early EDM paper by  \citeN{barnes:qmatrix} discusses the \emph{Q-matrix} approach, which maps test  item results  to latent  or  underlying knowledge structures. This mapping was originally performed using binary values, although these values can be straightforwardly mapped to probabilities if the binary values are replaced by a number ranging from zero to one. Thus, this approach constructs concept--question matrices that can be related to the performance of students using a variety of MF methods. 
Recent work by Desmariais and collaborators has made use of non-negative matrix factorization (NMF)  to extract Q-matrices from different data sets \cite{desmarais.beheshti.ea:item,desmarais:mapping,desmarais.naceur:matrix}, claiming NMF is far more interpretable than many MF techniques due to its insistence upon non-negative values in the two new matrices, which enables a probabilistic interpretation of the resulting matrices. 

The model of MF that is adopted very much affects the results that are obtained. In particular, the move from compensatory operations (which each added skill adds to the success of a topic) to more  conjunctive operators (where missing skills will lead to a student failing a test item) has been recognized \cite{barnes:novel,desmarais.beheshti.ea:item,desmarais:mapping}, but there is no clear consensus as to which factorization method should be used. Indeed, it is possible to factorize matrices describing student performance in other ways, and in Section~\ref{sec:matrixfactorization} one such alternative is presented.

While originally constructed by experts who defined the question to concept mappings, Q-matrices can be automatically constructed using simple hill-climbing algorithms which vary the number of concepts and the values in randomly seeded matrices, attempting to find a Q-matrix that best describes all student responses. In contrast to results obtained by traditional clustering methods, Q-matrices are more interpretable, which makes them interesting tools for communicating with both faculty and students about capabilities and weaknesses. The paper by \citeN{barnes:qmatrix} demonstrated that, in at least some cases, students who were given a self-guided option in an experiment were able to choose questions that were highly correlated with a Q-matrix ``least understood concept'' constructed from a simple lesson based tutorial. Furthermore, Barnes was able to demonstrate that a small sample of self-guided students who chose differently from the Q-matrix prediction ``could  have  benefited  from  reviewing  a  Q-matrix  selected concept'' before their final exam, stating correctly that a ``student  may  not  realize when he should review a particular topic''. Sometimes the items in which a student is most interested are not those from which they could best benefit. This suggests that RecSysTEL can perhaps be used to improve outcomes based upon profiles of student knowledge, particularly in more complex scenarios where student confusion is likely to increase. 

Q-matrices have been shown to compare favorably with Bayesian Knowledge Tracing (BKT) when it comes to predicting student success \cite{thai-nghe.horvath.ea:factorization}, but remain very difficult to use in scenarios based more around modeling student knowledge of topics. They tend to perform better when concepts and topics are distinct from one another, as happens with e.g. French and mathematics, but less well on trivia (or questions for which there is more overlap) \cite{winters:educational,desmarais:mapping}.

\subsection{Using Knowledge Gap Profiles in RecSysTEL}\label{sec:solution}

While many recommendation systems have been developed in TEL, they tend not to make use of MF for their profiling of students. Similarly, there appear to be few attempts to couple student profiles regarding knowledge with a scalable RecSys solution. Here, a full system is presented that: 
\begin{inparaenum}[(i)]
\item Takes note of student performance in a real world and open ended question answering scenario, 
\item constructs a learner profile based upon performance using MF that maps out their current knowledge gaps with respect to the  environment in which they are participating, and
\item recommends questions that will help them to remove this knowledge gap, while preferentially selecting questions that are similar to those that they have previously rated as interesting. 
\end{inparaenum}

\section{\name Techniques, Technologies, and Problem Definition}
\label{sec:rippleTechniques}

In this section more explicit details are provided on the MF algorithm and the Peer-Learning Environment that are used in this study. These details enable the definition of a tighter set of design requirements and the further refinement of the research problem in Section~\ref{sec:researchProblem}

\subsection{Matrix Factorization} \label{sec:matrixfactorization}	

Assuming $H_{N \times K}$  represents the latent factors underlying user behavior giving $h_u$, a vector of latent factors representing user $u$. Similarly,  $Q_{M \times K}$ is assumed to represent the latent factors of a question set, where $q_i$ is a vector of latent factors representing question $i$. After the mapping of users and questions to the latent factors, the rating of a user $u$ for a question $i$ can be approximated as:
\begin{equation}\label{eqn1}
	\hat{r}_{ui}= q_j^\mathsf{T} h_u  = \sum_{k=1}^K q_{ik}  h_{uk}
  \end{equation}
Matrix $\hat{R}=\{\hat{r}_{ui}\}$ is then used to capture all predicted ratings that users give a set of questions, with elements given by  Equation~\ref{eqn1}. 
The goal of MF is to learn the matrices $H$ and $Q$, which are used to compute values for $\hat{R}$, which approximate the unseen ratings that are actually given by users represented by $R$.
To learn these factors, a MF system minimizes the following regularized squared error term on the set of known ratings:
\begin{equation} \label{eqn2}
    \sum_{(u,i) \in R_{train}}(r_{ui}- q_i^\mathsf{T} h_u)^2  + \lambda(\norm{q_i}^2 + \norm{h_u}^2),
\end{equation}
where $(u,i) \in R_{train}$ represents $(u,i)$ pairs such that the rating of user $u$ for question $i$ is present in the training data set and $\lambda$ is a parameter controlling the extent of the regularization. 

The initial values of latent variables in $H$ and $Q$ are sampled from a standard normal distribution with zero mean and standard deviation of one. By performing stochastic gradient descent, in each iteration looping through the ratings in $R$, latent variables in $H$ and $Q$ are updated using Formulas~\eqref{eqn3} and~\eqref{eqn4} and tuned in order to locally minimize \eqref{eqn2}. The constant value $\gamma$ represents the learning rate, which is often determined using a validation set. 
\begin{align}
   h_u &= p_u + \gamma . ((r_{ui} - q_i^\mathsf{T} h_u ).q_i - \lambda.h_u)\label{eqn3}\\ 
  q_i &= q_i + \gamma . ((r_{ui} - q_i^\mathsf{T} h_u ).h_u - \lambda.q_i)\label{eqn4}
\end{align}
Extended research has aimed to improve this method generally. \citeN{Koren2008} illustrated that addition of mean normalization and a bias parameter for each user and item (in this case a question) can capture the effects associated with each, allowing only the true interaction portion of the ratings to be modeled in $H$ and $Q$. This method, referred to as {\bf Biased Matrix Factorization (BMF)}, is employed in \name. 

\subsection{The PeerWise Learning Environment}\label{sec:peerwise}	
PeerWise \cite{Denny2008} is a free web-based system in which students can both (i) create multiple-choice questions for sharing, and (ii) answer, rate, and discuss questions created by their peers. More than 1500 universities, schools and technical institutes from around the world have adopted Peerwise\footnote{\url{https://peerwise.cs.auckland.ac.nz/}}, and a number of papers have been published that discuss research completed using the platform  \cite{Hardy2014,Lumezanu2007,Bates2012,Purchase2010}.

In PeerWise students are expected to direct their questions towards the learning goals of the course.
Students receive immediate feedback on any answers that they record in the system. They are also shown a sample solution and data about how other students have answered the same question. This helps them to assess how well they are performing compared to their peers. Questions can also be  evaluated using peer-review, which encourages students to evaluate the quality and difficulty of any questions they answer, providing constructive open-ended comments in the process if appropriate. 
This feature enables asynchronous discussions over a period of time, where students can rate the quality of questions,  providing feedback for one another as to how they might be improved. 

The crowdsourcing process facilitated by this Peer-Learning Environment can lead to a repository of rich and high-quality multiple-choice questions that can be reused in future offerings of a course, as well as studied in their own right. 

PeerWise also includes several ``game-like" elements (such as badges, points and leaderboards) to inspire students to become more engaged with the platform. All activities remain anonymous to students; however, instructors are able to view the identity of question and comment authors, and to delete inappropriate questions. When students create a question, they can tag it with relevant topics, which can be student generated depending upon the settings chosen by an instructor. Instructors can also choose to predefine all tags to be used in the course if they feel that student generated tags will not work for their scenario. 

PeerWise currently does not provide personalized recommendations to students. However, the main PeerWise page where the questions are presented supports basic sorting functionality. Questions can be sorted based on different characteristics such as popularity, difficulty, and date of creation. A student can then manually search through the displayed questions to find suitable candidate questions for answering. Additional information about the reputation of the author and the number of times the question has previously been answered is also provided. 

\subsection{The Research Problem}
\label{sec:researchProblem}

The open ended structure of PeerWise leads to the specific research problem that this study aims to address: as a large and unstructured store of multiple choice questions, PeerWise can rapidly become un-navigable for students. This can lead to students focusing upon questions that reinforce existing knowledge, or satisfy their general interests, instead of those that are most likely to help them to satisfy study requirements. A RecSys could be used to discourage this behavior, but such a system must be able to both identify knowledge gaps in an individual learners profile, and find questions that are most likely to satisfy that knowledge gap. Ideally such a RecSys would be able to perform this function while prioritizing questions in which a user is likely to have an interest, as this will help to maintain their engagement with the system. 

The aim of this work is to enhance a RecSys using BMF with the concept of a learning profile, producing a RecSySTEL designed for the PeerWise learning environment. As a set of further requirements this tool should: enable a proof of concept scenario where users can choose different foci for the recommendations that they receive; support cold start users; scale appropriately while exhibiting robust behavior; allow users to understand the reason for the recommendations presented.

\section{Introducing \name}
\label{sec:riple}

\begin {figure}[h!]
\centering
\includegraphics[width=12 cm]{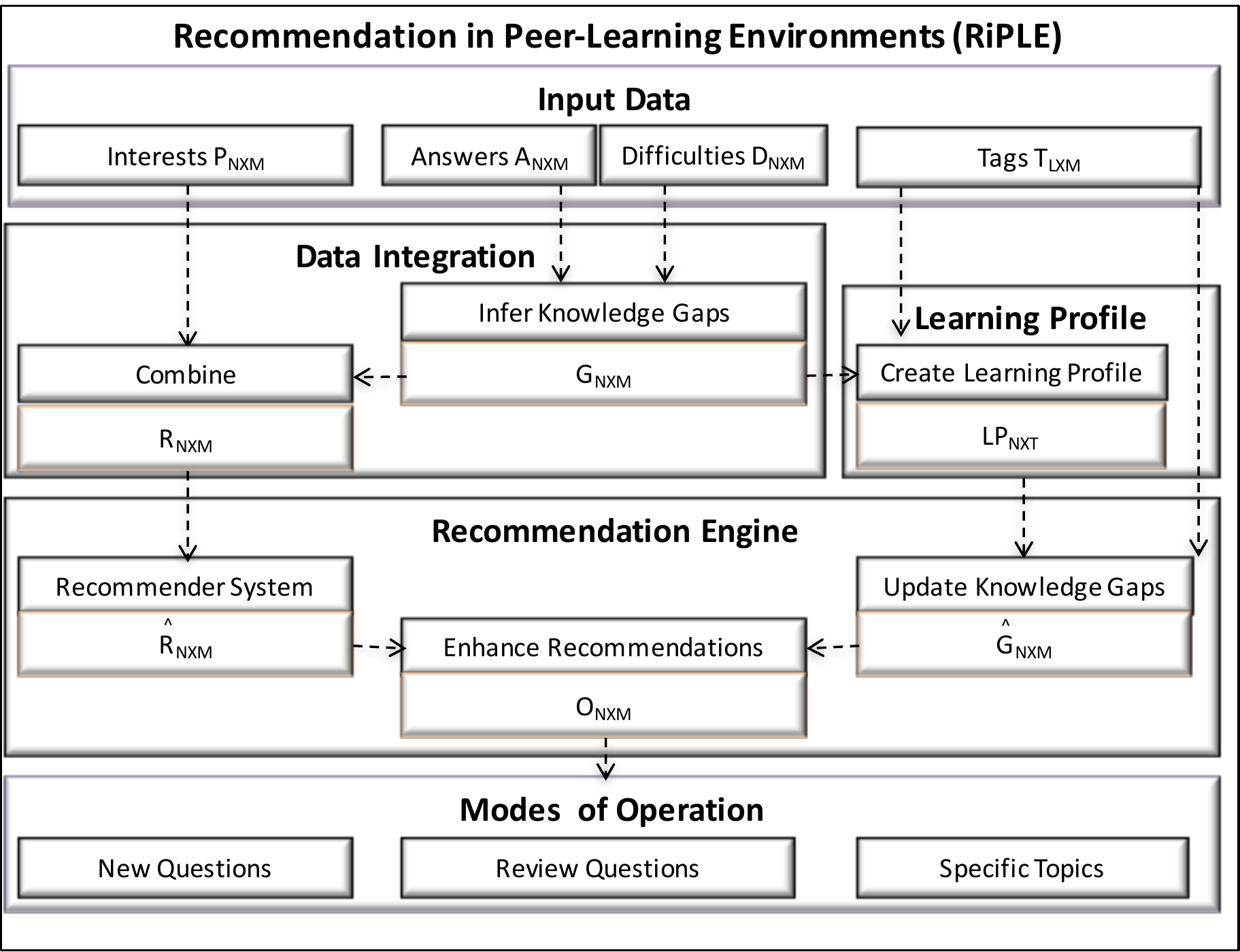}
\caption{\name: a framework for Recommendation in Peer-Learning Environments.\label{fig:overview}}
\end {figure}

At a high level, \name applies a suite of established approaches to harness data available in Peer-Learning Environments and provide personalized recommendations tailored towards each users' interests and knowledge gaps. \footnote{Source code for \name is available at \url{https://github.com/hkhosrav/RiPLE}}
\name is organized into five main modules: 
Input Data; Data Integration; Learning Profile; Recommendation Engine; and Modes of Operation. Figure~\ref{fig:overview} provides an overview of the system. Boxes in the Input Data module represent data gathered from PeerWise. The top part of the double boxes in the Data Integration, Learning Profile, and Recommendation Engine modules represent computations; the bottom part of the double boxes represent the results. The boxes in the Modes of Operation module represent the final selection and presentation of tailored recommendations to the users based on the operational mode selected. 
A summary of the notation used in describing \name  is presented in Table~\ref{tab:notation}. 

In what follows, Section~\ref{sec:input} provides more information about the input data, and Section~\ref{sec:integration} discusses how the data are used to infer knowledge gaps. Section~\ref{sec:profile} introduces the learning profile, Section~\ref{sec:recommendation} describes how learning profile enhanced recommendations are made, and Section~\ref{sec:modes} summarizes the different operational modes of the system.

\begin{table}[hbt]
\centering
\small{
\begin{tabular}{|l|l|l|l|}
\hline

\multicolumn{4}{|c|}{Input Data}  \\ 
\hline
$N$       & \multicolumn{3}{l|}{Number of users} \\
$M$       & \multicolumn{3}{l|}{Number of questions} \\
$L$       & \multicolumn{3}{l|}{Number of topics}        \\
$A_{N\times M}$       & \multicolumn{3}{l|}{Matrix, where $a_{ui}$ is 1 if user $u$ answers question $i$ correctly and 0 if answered incorrectly}   \\
$P_{N\times M}$       & \multicolumn{3}{l|}{Matrix, where $p_{ui}$ is the interest level user $u$ has expressed for question $i$}   \\
$D_{N\times M}$       & \multicolumn{3}{l|}{Matrix, where $d_{ui}$ is the difficulty level user $u$ has expressed for question $i$} \\                                                                              
$T_{M\times L}$       & \multicolumn{3}{l|}{Matrix, where $t_{ij}$ is  $\frac{1}{g}$ if $i$ is tagged with $g$ topics, including  $j$ and 0 otherwise} \\  \hline

\multicolumn{4}{|c|}{Data Integration}  \\ 
\hline
$G_{N\times M}$       & \multicolumn{3}{l|}{Matrix, where $g_{ui}$ is the knowledge gap of user $u$ based on question $i$}         \\ 
$\bar{d}_M$  & \multicolumn{3}{l|}{Vector, where $\bar{d}_i$ is the average difficulty expressed for question $i$ across all of the users} \\ 
$kgw$       & \multicolumn{3}{l|}{Constant to weight the relative impact of knowledge gaps and interests}   \\ 
$R_{N\times M}$       & \multicolumn{3}{l|}{Matrix, where $r_{ui}$ is the benefit that user $u$  would receive from doing  question $i$}   \\ 
\hline

\multicolumn{4}{|c|}{Learning Profile}                                                                                                                                 \\ \hline
$S_{N\times M}$       & \multicolumn{3}{l|}{Matrix, where $s_{ui}$ is 1 if user $u$ has attempted question $i$ and 0 otherwise}       \\  
$C_{N\times L}$       & \multicolumn{3}{l|}{Matrix, where $c_{uj}$ is the weighted sum of questions tagged with topic $j$ attempted by user $u$} \\  
$LP_{N\times L}$       & \multicolumn{3}{l|}{Matrix, where $lp_{uj}$ is the approximated knowledge gap of user $u$ on topic $j$}    \\  
$\bar{lp}_L$  & \multicolumn{3}{l|}{Vector, where $\bar{lp}_j$ is the average knowledge gap for topic $j$ across all of the users} \\ 

$\beta$  & \multicolumn{3}{l|}{constant parameter controlling the impact of the learning profile} \\  \hline

\multicolumn{4}{|c|}{Recommendation Engine}          \\  \hline
$O_{N\times M}$       & \multicolumn{3}{l|}{Matrix, where $o_{ui}$ is the predicted personalized score of question $i$ for user $u$}                   \\    
$\hat{R}_{N\times M}$       & \multicolumn{3}{l|}{Matrix, where $\hat{r}_{ui}$ is the predicted benefit that user $u$  would receive from doing  question $i$}      \\ 
$\bar{r}_M$  & \multicolumn{3}{l|}{Vector, where $\bar{r}_i$ is the average benefit of question $i$ across all of the users} \\ 
$\hat{G}_{N\times M}$ & \multicolumn{3}{l|}{Matrix, where $\hat{g_{ui}}$ is the inferred knowledge gap of user $u$ on question $i$ based on $lp_{u}$}            \\ \hline

\end{tabular}
}
\caption{A summary of the notation used in describing \name}
\label{tab:notation}
\end{table}

\subsection{Input Data} 
\label{sec:input}

As discussed in Section~\ref{sec:peerwise}, the input data consists of multiple-choice questions that are tagged with distinct topics and user ratings for quality and difficulty. Table~\ref{tab:notation} summarizes all input data, characterizing it with $N$ to denote the number of users that are registered on the Peer-Learning Environment, $M$ for the number of multiple-choice questions that have been contributed to the environment, and $L$ as the number of distinct topics that have been used to tag the questions. The expressed user ratings  and correctness values in a PeerWise data set are used to populate the matrices. The interests ($P$) are originally stored in PeerWise as integers in the range of [0, 5] and the difficulties ($D$) as integers in the range of [0, 2]. 
Data input to \name from PeerWise are organized in four matrices:
\begin{compactdesc}
\item[Interests, $P$:] Each user can rate the quality of the questions that they have answered. This information is represented in a  matrix $P_{N \times M}$, where  $p_{ui}$ captures the interest level user $u$ has expressed for question $i$. These ratings are stored  as a value between between 0 and 1 when expressed and as $Null$ otherwise.
\item[Difficulties,  $D$:] Each user can rate the difficulty of the questions that they have answered. This information is represented in a  matrix $D_{N \times M}$, in which $d_{ui}$ captures the difficulty level user $u$ has expressed for question $i$. These ratings are stored  as a value between between 0 and 1 when expressed and as $Null$ otherwise.
\item[Tags, $T$:] Each question can have 0 to $L$ topics assigned (i.e. tagged) to it.  The information on topics assigned to each question is represented in a matrix $T_{M \times L}$, in which $t_{ij}=0$  indicates that question $i$ is not tagged with topic $j$ and $t_{ij}=\frac{1}{g}$ indicates that question $i$ is tagged with $1 \leq g  \leq L$ associated topics, including  $j$.
\item[Answers, $A$:] The correctness of the answers provided by the users is represented in a matrix $A_{N \times M}$. If a user $u$ answers a question $i$ correctly then the matrix entry is set at $a_{ui}=1$,  $a_{ui}=0$ indicates an incorrect answer, and $a_{ui}= Null$ indicates that question $i$ has not been attempted by user $u$.
\end{compactdesc}

\subsection{Data Integration} \label{sec:integration}

This module uses the input data $A$, $D$, and $P$ to produce an overall rating matrix $R_{N \times M}$, where $r_{ui}$ captures the extent to which a user $u$ would benefit from answering a question $i$. This matrix represents how much the knowledge gaps of individual users can be reduced while keeping engagement at a maximum.  $R_{N \times M}$ is constructed in two steps.

\paragraph{Inferring Knowledge Gaps} First,  information from $D$ and $A$ is combined to create a scoring function that maps user performance to knowledge gaps. The function determines user $u$'s  lack of knowledge about question $i$, independent of their performance on other questions. Matrix $D$ is used for computing a vector $\bar{d}$, where $\bar{d}_i = \frac{\sum_{u=1}^N{d_{ui}}}{N}$ represents the average rating for question $i$ across all of the users. 
The scoring function produces a matrix $G_{N\times M}$, where $g_{ui}$ infers user $u$'s lack of knowledge about question $i$.    
\begin{equation} \label{eq:score}
g_{ui} = (1-a_{ui}) (\frac{0.5 - a_{ui}}{1 + \stackrel{-}{d_i}}) +  a_{ui}(\frac{0.5 - a_{ui}}{2 - \stackrel{-}{d_i}})
\end{equation}
A higher value for \eqref{eq:score} indicates a larger knowledge gap. This function captures two intuitions about user responses: 
\begin{enumerate}
\item An incorrect answer indicates a knowledge gap for topics related to that question. The significance of the gap may be approximated by the difficulty level of the question; answering an easy question incorrectly suggests a large gap and answering a hard question incorrectly suggests a smaller  one.
\item Answering questions correctly provides evidence that no knowledge gap exists, suggesting user competency on the related topics.  The significance of the competency may be approximated by the difficulty level of the question; answering an easy question correctly illustrates low level of competency and answering a hard question correctly illustrates a higher level of competency.
\end{enumerate}
Equation~\eqref{eq:score} uses summation to combine these intuitions. The first part of the equation is positive, indicating a knowledge gap for an incorrectly answered question $i$  weighted by $\hat{d}_i$. The second part contributes to the score with a negative value, indicating competencies, when the question is answered correctly. For example, answering an easy question $i$ with $\bar{d}_i=0.1$ incorrectly results in the scoring function returning 0.45, and answering a hard question $i'$ with $\bar{d}_{i'}=0.7$ correctly results in the scoring function returning -0.38. Given that the difficulties are stored as values between 0 and 1, values in $G$ remain in the range of [-0.5, 0.5].

\paragraph{Combining Knowledge Gaps and User Interests} The knowledge gaps inferred from the previous calculation are then combined with students' interests to produce a matrix $R$, which captures the extent to which users would benefit from answering different questions:
\begin{equation} \label{combine}
R =   kgw G + (1-kgw)P.
\end{equation}

Here, a weight term ($0 \leq kgw \leq 1$) is used to represent the impact of the knowledge gaps upon users. This may be be set as default for an entire cohort, or for individual students. Both students and  instructors could set $kgw$ at the individual level, or  additional machine learning techniques could also be used in future work. The value of $kgw$ allows the weight of the knowledge gaps and the interests of the students to be adjusted along a spectrum between what the users of \name need (to master course content) and what they prefer (to improve engagement).   


\subsection{Learning Profile} \label{sec:profile}

This module uses the input data $T$ and the knowledge gap matrix $G$ provided by Data Integration module to produce a student-topic learning profile $LP_{N \times L}$, in which each vector ($lp_u$) approximates a user's knowledge gaps across all of the topics associated with the course. A negative value in the vector, i.e. $lp_{uj}<0$ indicates that the user $u$ has demonstrated some knowledge on topic $j$, a positive value indicates a knowledge gap on that topic, and 0 represents a neutral state, where the  positive and negative scores 
have balanced each other out for that particular topic. 
The learning profile is computed in two steps:
\begin{compactenum}
\item Matrix $G_{N \times M}$ stores information about the lack of knowledge exhibited by all users for each question, and matrix $T_{M \times L}$ stores information about the tags associated with each question. 
Multiplying the two 
($G T$)
allows for an understanding to be gained about topic-level knowledge gaps in the system \emph{per se}. 
The value stored in cell $[u, j]$  
of the resulting matrix depends on the number and weight of questions tagged with topic $j$ that have been attempted by each user $u$. This means that the values in this matrix require normalization. 
\item Normalization is achieved using a user-topic count matrix  $C_{N \times L}$, in which $c_{uj}$ represents the weighted sum of questions attempted by user $u$ that have been tagged with topic $j$. This matrix can be computed using
$C= S  T$,
in which $s_{ui}$ is 1 if question $i$ is attempted by user $u$ and 0 otherwise. 
\end{compactenum}
Putting the two steps together, the learning profile is computed using the following formula:

\begin{equation} \label{lp}
LP =  \frac{G T}{C}
\end{equation}
This learning profile may be shared with students to inform them of their knowledge gaps and competencies at a topic-level, enabling them to understand the reason for \name's recommendations. As mentioned in Section~\ref{sec:researchProblem}, this is one of the core requirements of the system. It would also allow them to compare their performance with the cohort. Such learner centered learning analytics could help to stimulate self-reflection among students, as well as providing an early alert for those that are performing below their targeted performance goal. Future work will seek to explore this intriguing possibility. 

\paragraph{Cold-start users}
A user that has answered zero or very few questions is referred to as a cold-start user. The knowledge gaps of a cold-start user $c$ that has answered zero questions are represented with a vector of zeros. In this scenario the system is unable to reliably infer user c's knowledge gaps, and therefore, cannot make meaningful recommendations.To address this issue, the knowledge gaps for cold-start users are estimated using the average knowledge gaps for the cohort, a vector $\bar{lp}$, where $\bar{lp}_j=\frac{\sum_{u=1}^N{lp_{uj}}}{N}$ represents the average knowledge gap for topic $j$ across all of the users in $LP$.  This solution to the cold start problem makes the assumption that most new users will have similar knowledge gaps to the average user, an assumption that could be questioned, but which should result in better initial recommendations for the majority of users in the system (by definition).


\subsection{Recommendation Engine} \label{sec:recommendation}
This module uses the benefit matrix $R$ (produced by the Data Integration module) and learning profile matrix $LP$ (produced by the Learning Profile module) to produce a matrix $O$ which contains vectors $o_u$ predicting the extent to which user $u$ would benefit from each of the questions in the PeerWise system. The process of making these recommendations is accomplished in three steps. Again, cold start users require unique processing for this module (see below).

\paragraph{Exemplary Recommendation}
First, matrix factorization as described in Section~\ref{sec:matrixfactorization} is employed to characterize users and questions using vectors of latent factors that form $\hat{R}_{N\times M}$. This  predicts the extent to which users might benefit from completing unseen questions. 

\paragraph{Updating the student-question knowledge gap} This step uses the matrices $LP$ (produced by the Learning Profile module) and the input tag matrix {T} to produce an updated student-question matrix $\hat{G}_{N\times M}$, in which $\hat{g}_{ui}$ approximates user $u$'s knowledge gap of question $i$ based on $lp_u$ and the tags associated with $i$. This is accomplished using the following equation: 

\begin{equation} \label{ghat}
\hat{G} =   {LP} T^\mathsf{T}
\end{equation}

Multiplying $LP_{N \times L}$ and $T^\mathsf{T}_{L \times M}$  propagates the lack of knowledge from course topics over to the associated questions.

\paragraph{Enhancing Recommendations} The updated benefit matrix $\hat{R}$ and the updated student-question matrix $\hat{G}$ that were extracted in the previous two steps are used to create the recommendation output matrix {O}, in which $o_{ui}$ represents the personalized rating of question $i$ for user $u$ tailored towards their knowledge gaps and interests. Values in $O$ are computed using the following formula: 

\begin{equation} \label{output}
O = \hat{R} + \beta \hat{G}
\end{equation}
where $\beta$ is a parameter controlling the impact of the learning profile, which may be determined using a validation set. 

\paragraph{Cold-start Users} The regularized squared error used in matrix factorization sets the latent factors of a user $u$ based on two terms: minimizing the first term tunes the latent factors of $u$ for predicting the ratings in the training set and minimizing the second term helps keep the latent factors small to avoid over-fitting (see Equation~\eqref{eqn2}). Since cold-start users do not have any ratings in the training set, the first term does not affect the outcome, so the learning algorithm is encouraged to reduce the error rate of the cost function by setting the latent factors all to zero without paying a penalty on the first term. Since multiplying a vector of zeros by the latent factors of any question returns zero, the system is unable to make any meaningful recommendation for cold-start users. 

One possible solution for overcoming this problem is to use mean normalization. Let $\bar{r}$ be a vector storing the average rating for each question, so $\bar{r}_i$ represents the average rating for question $i$ across all of the users in $R$. During the learning phase, values in $R$  are normalized with $\bar{r}$  using the following formula:
\begin{equation} \label{adjustcsrhat}
r_{ui}= r_{ui} - \bar{r}_i.
\end{equation}
With this update, after the learning phase, values in $R$ have the following interpretation:  $r_{ui} > 0$ indicates that $u$ would rate $i$ higher than average,  $r_{ui} < 0$ indicates that $u$ would rate $i$ lower than average, and $r_{ui} \simeq 0$ indicates that $u$ would rate $i$ close to the average. Using mean normalization has the benefit that the system's ratings for a cold-start user $c$, which is $r_c= \{0\}$,  has now the interpretation that $c$'s rating of each of the questions is the global average for that question. 

After the learning phase, values are de-normalized and stored back in $\hat{R}$ using the following formula, in which $\bar{r}_i$ is added back to the ratings for question $i$
\begin{equation} \label{restorecsrhat}
\hat{r}_{ui}= r_{ui}  + \bar{r}_i.    
\end{equation}

\subsection{Modes of Operation} \label{sec:modes}

The system operates in three different modes, each having its own advantages and use case. The modes select questions to present to the user $u$ from their vector in the output matrix $O_u$ (produced by the Recommendation Engine).
\begin{compactdesc}

\item [Exploring new questions:] In this mode, the system is designed to present users with questions that they have not seen before, preferentially choosing the unseen questions with the highest recommendation values for the user $u$, in the vector $O_u$. This mode is ideal for general practice, allowing users to explore new questions that are tailored towards their interests and reducing their knowledge gaps.
\item [Reviewing answered questions:] In this mode, the system is designed to present users with questions that they have seen before, preferentially choosing the seen (answered) questions with the highest recommendation values for the user $u$, in the vector $O_u$. This mode is ideal for preparation for exams; the system prioritizes questions that cover topics where the user lacks knowledge and topics in which they are interested. 
\item [Focusing on specific topics:] In this mode, the system is designed to present the user with questions from selected topic(s),  regardless of whether they were previously attempted or not, choosing those which have the highest recommendation values for the user $u$, in the vector $O_u$. 
This mode is ideal for practice on specific topics, in which the system prioritizes questions that the user finds most interesting and helpful in reducing their knowledge gaps.
\end{compactdesc}

\section{Simple Example}
\label{sec:example}

\begin {figure}[h]
\centering
\includegraphics[height=13cm]{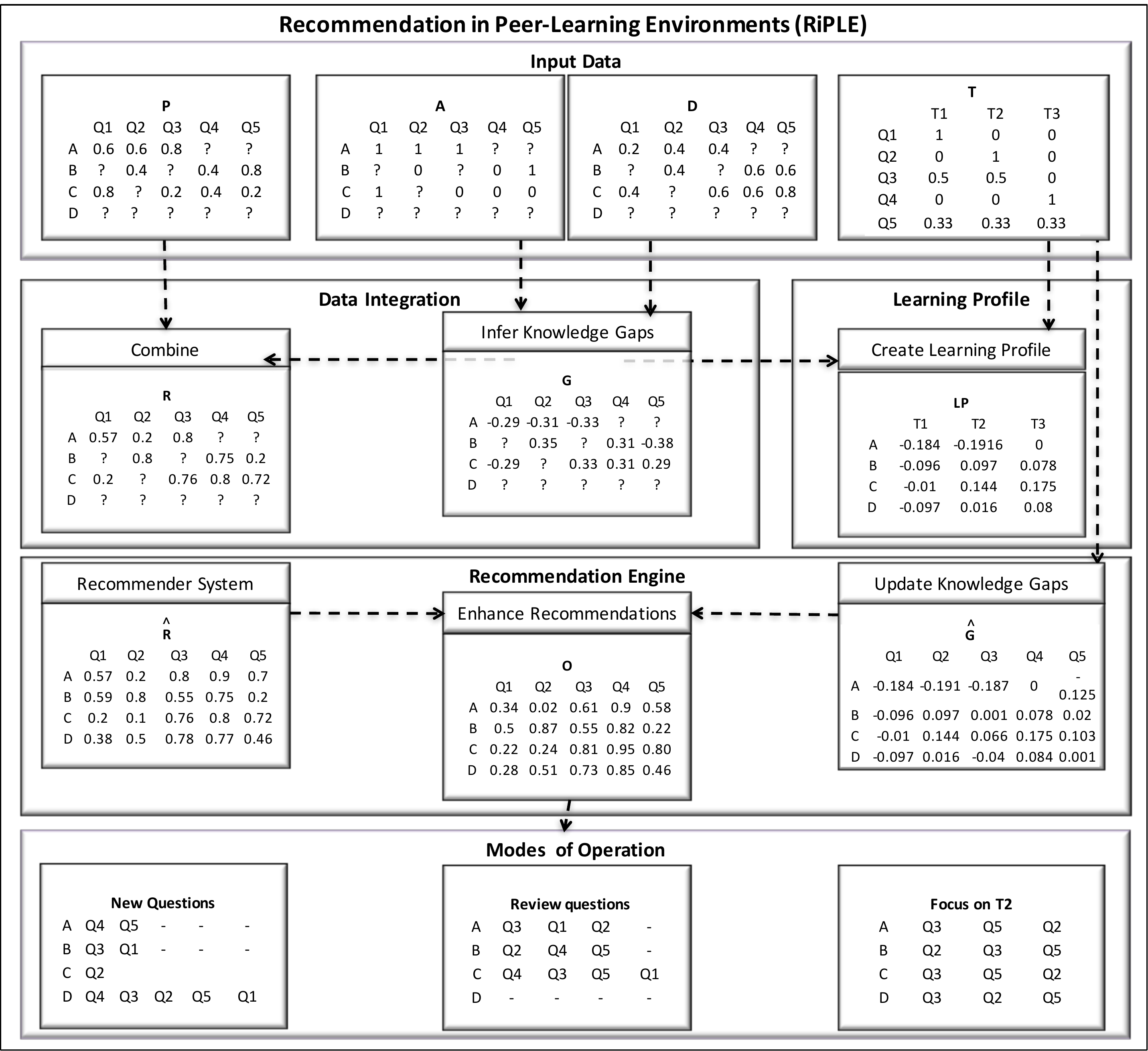}
\caption{ An example of \name with four students, five questions, and three topics that shows how the system operates. \label{fig:example}}
\end {figure}

In order to ground the above discussion of \name, a simplified example with four students, five questions and three topics is presented. Figure~\ref{fig:example} shows an overview of the example based on the framework provided in Figure~\ref{fig:overview}. In this example, Alice (A), Bob (B), and Catherine (C) are all defined as existing active users and Dean (D) is a new cold-start user of \name. For this simple example the two parameters are set to $\beta=1$ and $kgw=0.8$.

In the current scenario, Alice has correctly answered the first three questions, and she has not found them to be very challenging. Because of her correct answers on topics  $T1$ and $T2$, her learning profile vector [-0.18, -0.19,  0] indicates a significant lack of knowledge gap on those topics, and a neutral state on $T3$ since she has not attempted any questions on that topic. Forming $\hat{G}$ by propagating the information from the knowledge gaps over to questions based on the associated tags leads to the indication that answering $Q4$ followed by $Q5$, which both focus on $T3$ benefits her the most in terms of reducing her knowledge gaps.  Considering the output vector from matrix $O$ for Alice, in the explore mode \name recommends answering $Q4$ over $Q5$ since it would help her the most in overcoming the existing knowledge gap on $T3$. In the review mode, the system recommends answering $Q3$ over the other questions since they all contribute roughly the same in overcoming her knowledge gaps, but she has expressed a higher interest towards $Q3$. In the focus mode assuming $T2$ is selected, the systems recommends $Q3$, with a slight edge, over Q5 because of her high interest on that question. Assigning a larger value to $\beta$ (e.g. 1.5 instead of 1) would have resulted in $Q5$ being recommended over $Q2$.

Bob has answered three questions altogether, two of  which are answered incorrectly. Matrix $D$ also shows that he has rated the questions as more challenging than Alice. Because of his incorrect answers on topics  $T2$ and $T3$, his learning profile vector [-0.096,  0.097, 0.078]  indicates a knowledge gap on those topics. The predicted gap for topic $T2$ is greater since the question answered incorrectly on $T2$ had a lower level average difficulty compared to the question that was answered on $T3$. Bob answered an easy question correctly on $T1$, so the vector shows a slight competency on that topic. Forming $\hat{G}$ leads to the prediction that answering $Q2$, focused on $T2$, followed by $Q4$, focused on $T3$, would benefit Bob the most in terms of reducing his knowledge gaps. Considering the output vector from matrix $O$ for Bob, in the explore mode the system redirects recommendations from $\hat{R}$ to recommend answering $Q3$ over $Q1$, allowing him to receive further practice on $T2$.  In both the review mode and the focus mode, assuming $T2$ is selected, the system recommends answering $Q2$ since this would help overcome knowledge gaps on $T2$ and Bob has expressed relatively high interest in that question.

Catherine (C) has answered four questions, but only one of them correctly.  Based on her two incorrect answers on $T3$, one incorrect answer on $T2$, and one correct and one incorrect answer on $T1$ her learning profile vector is computed as [-0.01,  0.144,  0.175]. The slight competency on $T1$ arises because the $T1$  weight of the question she answered correctly ($Q1$) was greater than the one she answered incorrectly ($Q5$). Consistent with these findings, $\hat{G}$ indicates that she would benefit the most from answering $Q4$, which solely focuses on $T3$. Considering the output vector from matrix $O$ for Catherine, in the explore mode the system recommends answering $Q2$ since it is the only unanswered question. In the review mode, the system recommends answering $Q4$, helping her overcome knowledge gaps on $T3$. Despite her high interest for $Q1$, it receives a very low overall rating as it does not help her overcome her major knowledge gaps. Assigning a smaller value to $kgw$ (0.1 instead of 0.8) would have resulted in $Q1$ being recommended over $Q4$. In the focus mode, assuming $T2$ is selected, the system recommends $Q3$ with a slight edge over $Q5$ because of the higher rating in $\hat{R}$.

Dean has not answered any questions, so he is a cold-start user. Since mean normalization is used, Dean's latent factors in $\hat{R}$ are filled with average ratings from the training data set (e.g.,  $\hat{r}_{DQ1}$ is equal to $0.38$, which is the average of $r_{AQ1}$ and $r_{CQ1}$). By using mean normalization, the system approximates Dean's knowledge gaps based on the knowledge gaps of the cohort, $\bar{lp}$; therefore,  he is expected to benefit the most from questions on $T3$ and the least from questions on $T1$. Considering the output vector from matrix $O$ for Dean, in the explore mode the system recommends answering $Q4$, which is also recommended to many regular users. In the review mode, the system cannot make any recommendations since he hasn't answered any questions before. In the focus mode, the questions from each topic that the cohort would mostly benefit from would also be recommended to Dean. 

\section{Validation: Using Synthetic Data Sets}
\label{sec:validation}

The goal of \name is to provide accurate recommendations that help users to overcome knowledge gaps, while keeping their engagement to a maximum by prioritizing questions that are of interest to them. In this section, the behavior of the system is validated and examined under different circumstances using synthetic data sets, in which the underlying knowledge gaps of the students are pre-defined. Each experiment is repeated five times. Reported values are the average results across the five runs.  In these experiments users that have answered less than three questions are considered cold-start users.

The following metrics are used for evaluating the output.
\begin{compactdesc}
\item[Metric for Question-level recommendations] As used commonly in recommender systems evaluations,  Root Mean Squared Error (RMSE) is used for measuring the error in the recommendation:
\begin{equation} \label{rmse}
RMSE =  \sqrt{\frac{ \sum_{(u,i) \in ds}(r_{ui}-\hat{r}_{rui})^2}{|ds|}} 
\end{equation}
where $ ds$ is the set of all pairs of (u, i) in the data set that RMSE is being reported on.
\item[Metric for topic-level recommendations] Accuracy of the model in terms of recommending questions that match students' most significant knowledge gap:
\begin{equation} \label{accuracy}
Accuracy = \frac{match}{|ds|}
\end{equation}
where $match$ is the number of instances $\in ds$ where the topic of the recommendation matches student's most significant knowledge gap.  
\end{compactdesc}

\subsection{Template for Generating Synthetic Data Sets}
The experiments discussed in this section make use of synthetic data sets generated using the following sequence of steps. First, a set of users with pre-defined knowledge gaps over a set of topics are created. Second, a set of questions with a pre-defined topic, level of difficulty and discrimination is generated. Knowledge gaps must sum to one, and are constructed by sampling from a Dirichlet distribution, where $\alpha$ defines the sparsity of the distribution; a smaller value of $\alpha$ creates a sparser distribution over knowledge gaps, producing synthetic users with a large gap over one topic. The topics associated with a question are sampled from a discrete uniform distribution; their level of difficulty and discrimination are both sampled from a normal distribution.
The probability of a user $u$ answering a question $i$ correctly is computed using a 2 parameter logistic Latent Trait Models (IRT) model from classical Item Response Theory \cite{Drasgow1990}, as recommended by \cite{Desmarais2010}:
\begin{equation} \label{responseTheory}
 \frac{1}{1+e^{-a_i(\theta_s - b_i)}}
\end{equation}
where $\theta_s$ represents user's average lack of knowledge gaps (competencies) in the topic(s) associated with question $i$ , $b_i$ is the difficulty level and $a_i$ is the discrimination level of question $i$. The difficulty level user $u$ has expressed towards question $i$ is sampled from a normal distribution based on the difficulty level of $i$. The interest level that user $u$ has expressed towards question $i$ is sampled from a uniform distribution.

In all generated data sets 400 users, 1100 questions, and 22000 answers are sampled, which roughly matches the numbers from the historical data set that is used for exploration in Section~\ref{sec:exploration} If not otherwise stated, the hyper-parameters are set using the following default values: $\alpha = 0.1$, $L = 10$, $\beta = 0.1$, $kgw = 0.8$, $\gamma = 0.1$, $k = 5$. Results are evaluated using 5-fold cross-validation.


\subsection{Impact of Varying Parameters in Synthetic Data Set Generation}

\paragraph{Impact of the Sparsity of the Pre-defined Knowledge Gaps ($\alpha$):}

\begin{figure*}[h!]
    \centering
    \begin{subfigure}[t]{0.5\textwidth}
        \centering
        \includegraphics[height=1.5in,width=2.7in]{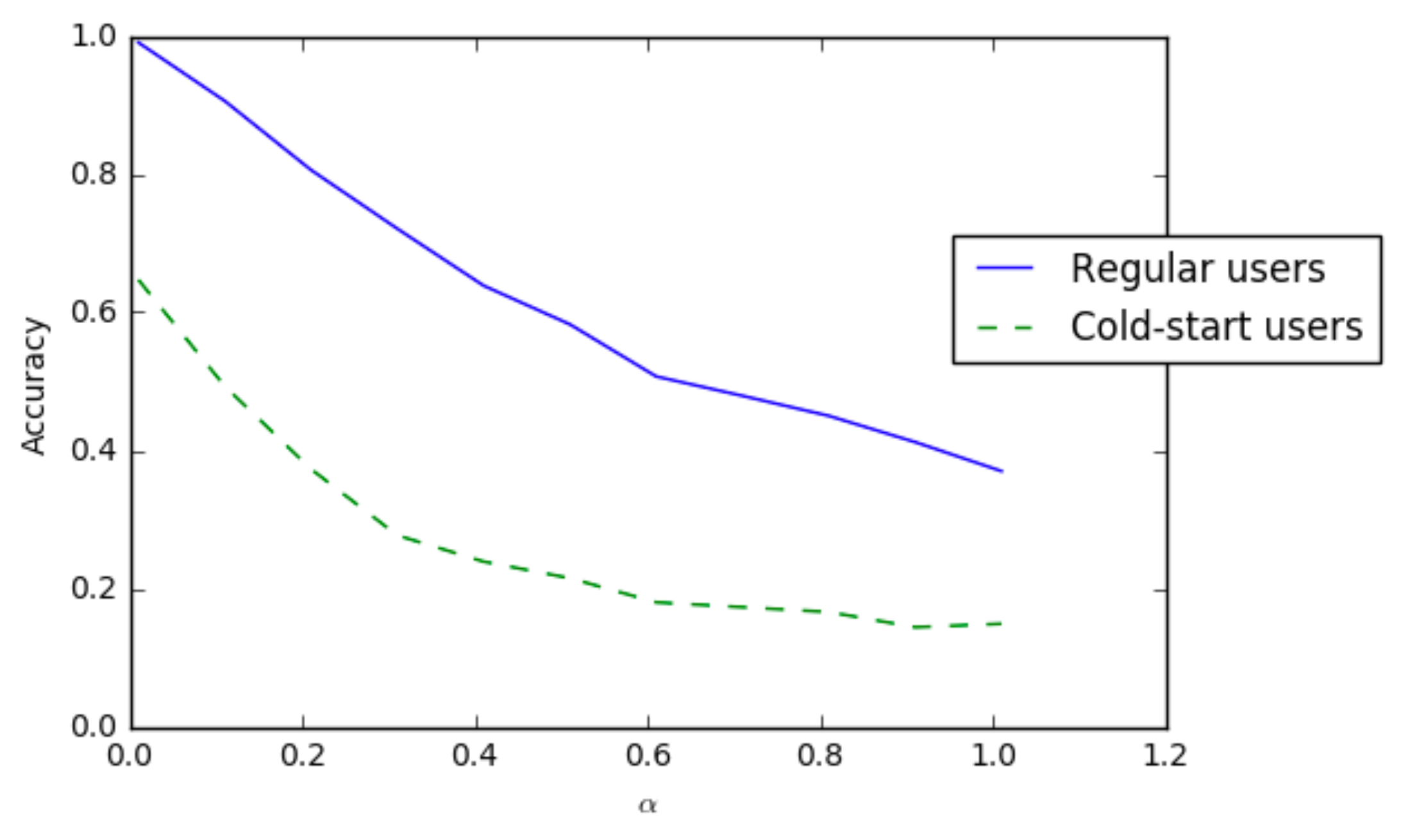}
        \caption{Accuracy as $\alpha$ is increased }
    \end{subfigure}%
    ~ 
    \begin{subfigure}[t]{0.5\textwidth}
        \centering
        \includegraphics[height=1.5in,width=2.7in]{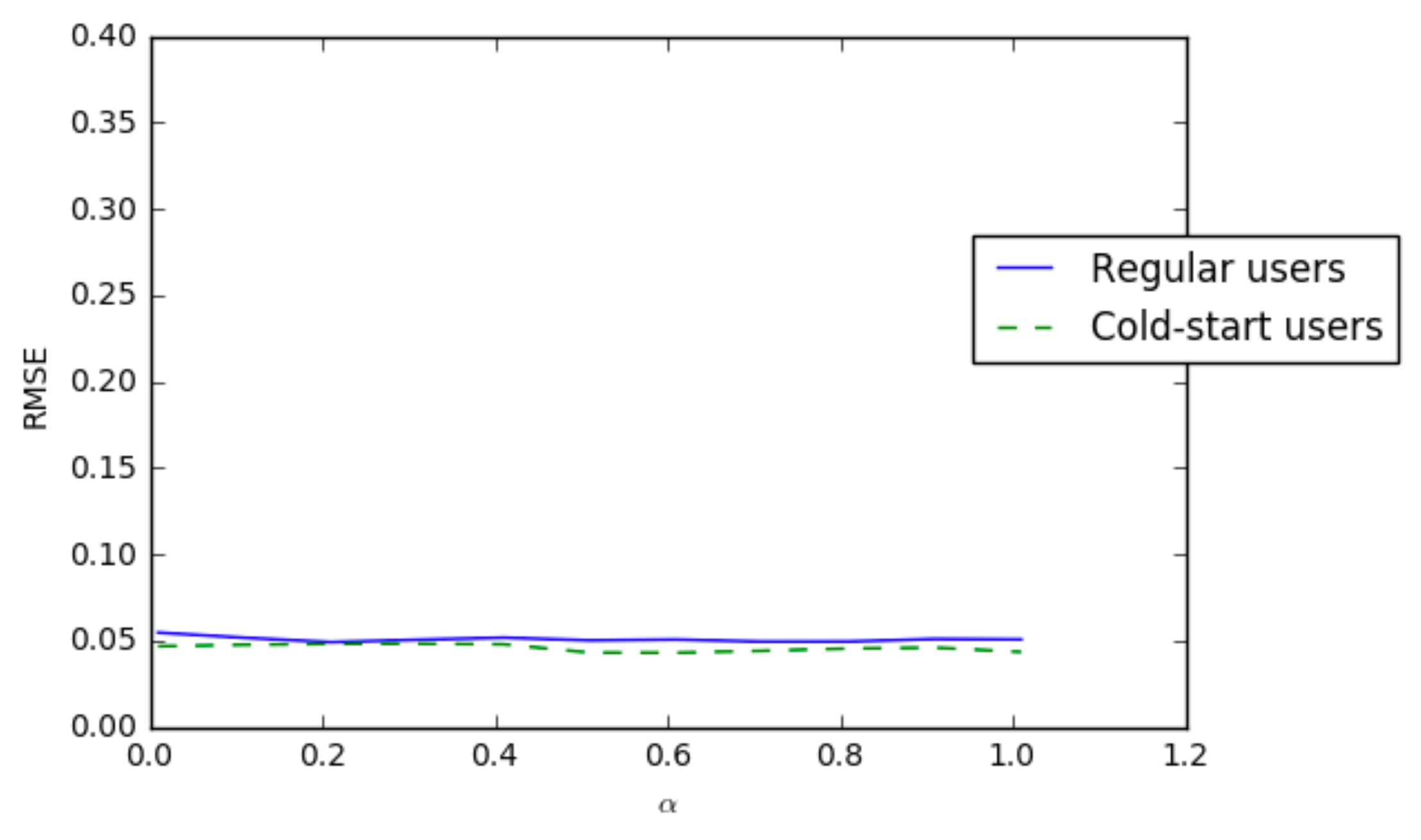}
        \caption{ RMSE as $\alpha$ is increased }
    \end{subfigure}
    \caption{Changes in accuracy and RMSE as the sparsity of the pre-defined knowledge gaps is decreased. \label{fig:alpha}}
\end{figure*}
Figure~\ref{fig:alpha} illustrates the effect of $\alpha$, which defines the sparsity of user knowledge gaps among the topics, upon the accuracy and RMSE of \name.
For regular users, \name is able to provide recommendations that target users most significant knowledge gaps when $\alpha$ is small. Increasing  $\alpha$, leads to the simulation of users with less extreme pre-defined knowledge gaps, making it more challenging for the system to accurately identify their most significant gap. For cold-start users, the system's accuracy is lower, as expected. We note that \name is still able to provide reasonable recommendations, considering the limited data available on those users, as long as $\alpha$ does not grow too large. Since users' knowledge gaps are defined as a vector that sums to one, changes in $\alpha$ do not have a significant impact on the overall probability of a user answering questions correctly, but only moves the knowledge gaps among topics, therefore the RMSE remains quite stable as $\alpha$ is increased.

\begin{figure*}[h]
    \centering
    \begin{subfigure}[t]{0.5\textwidth}
        \centering
        \includegraphics [height=1.5in,width=2.7in]{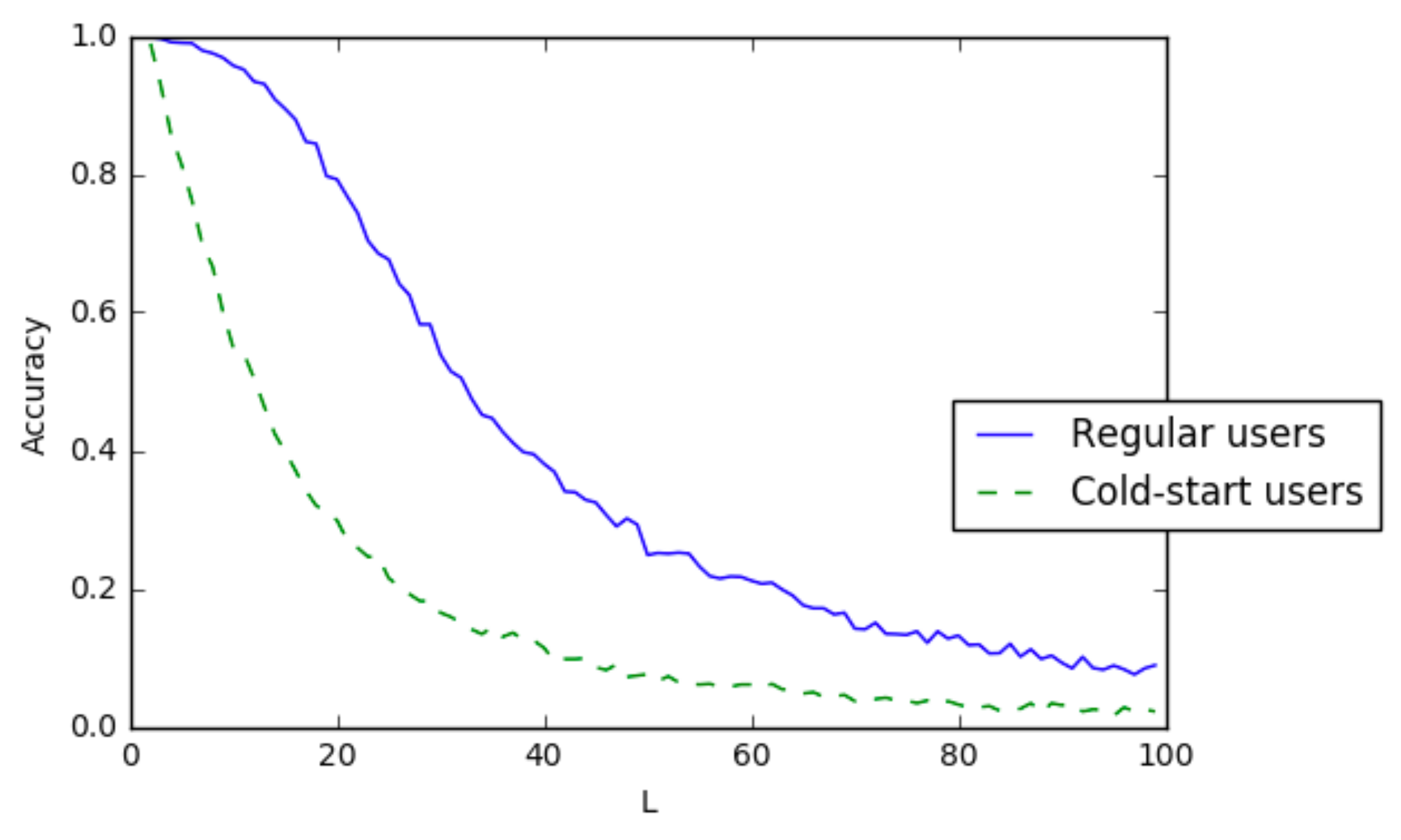}
        \caption{Accuracy as $L$ is increased}
    \end{subfigure}%
    ~ 
    \begin{subfigure}[t]{0.5\textwidth}
        \centering
        \includegraphics[height=1.5in,width=2.7in]{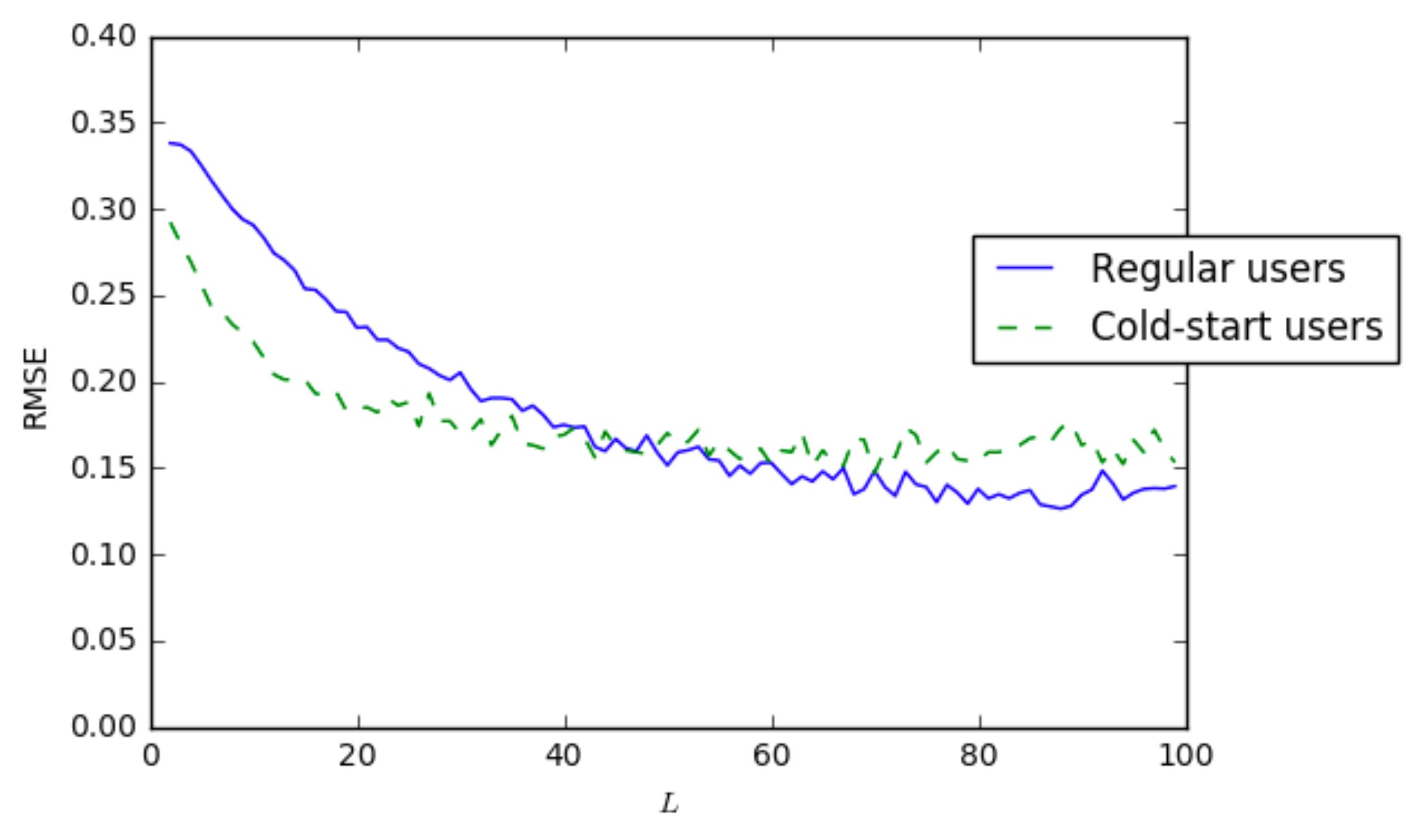}
        \caption{ RMSE as $L$ is increased }
    \end{subfigure}
    \caption{Changes in accuracy and RMSE as the number of topics is increased.\label{fig:topic}}
\end{figure*}

\paragraph{Impact of Number of Topics ($L$):}
Figure~\ref{fig:topic} illustrates the impact of increasing $L$, which shows the distinct number of topics that have been used for tagging the questions.

For regular users with $L<10$, \name is able to provide recommendations targeting their most significant knowledge gap over 90\% of the time.  For $10 \leq L \leq  20$, which would be the case for most of the commonly taught courses, the system remains relatively reliable being able to identify the most significant gap over 80\% of the time. For more extreme cases where $20 \leq L \leq  100$, the system does computationally scale; however, the task becomes much more challenging, and the accuracy drops significantly. For cold-start users, the system finds it more challenging to determine the most significant knowledge gap of a user (as expected) since they are unlikely to have encountered a question on that topic. 

As described in \eqref{responseTheory}, the probability of a user answering a question correctly relies significantly on the pre-defined knowledge gaps of the user. When dealing with users with sparse knowledge gaps, the sampled values determining whether or not a user correctly answers a question with an unknown topic has
the highest standard deviation, unpredictability, in the case of $L=2$ since there is approximately a 50-50 split between the question being answered correctly and incorrectly; therefore, 
the highest RMSE is observed when $L=2$. For $L>2$, the standard deviation of sampled values is reduced since most of the questions would have a higher probability of being answered correctly.

\paragraph{Summary} The results presented in this section have demonstrated that both $\alpha$ and $L$  have a significant impact on the performance of \name.  For small values of $\alpha$ and $L$, which contribute to the creation of a simple environment with strong correlations among values in the data set, \name is able to provide recommendations that strongly match users' knowledge gaps, validating the theoretical foundations of the system.  Larger values of $\alpha$ and $L$ can be used to create more realistic data sets, in which \name is still able to perform relatively well. Extreme values of $\alpha$ and $L$ (that lead to the creation of data sets more complicated than expected in real data sets on Peer-Learning Environments)  demonstrate the scalability of the system, showing that it exhibits robust behavior under more extreme circumstances.

\subsection{Impact of Varying \name Model Parameters}

\paragraph{Impact of the learning profile ($\beta$):}

\begin{figure*}[h]
    \centering
    \begin{subfigure}[t]{0.5\textwidth}
        \centering
        \includegraphics[height=1.5in,width=2.7in]{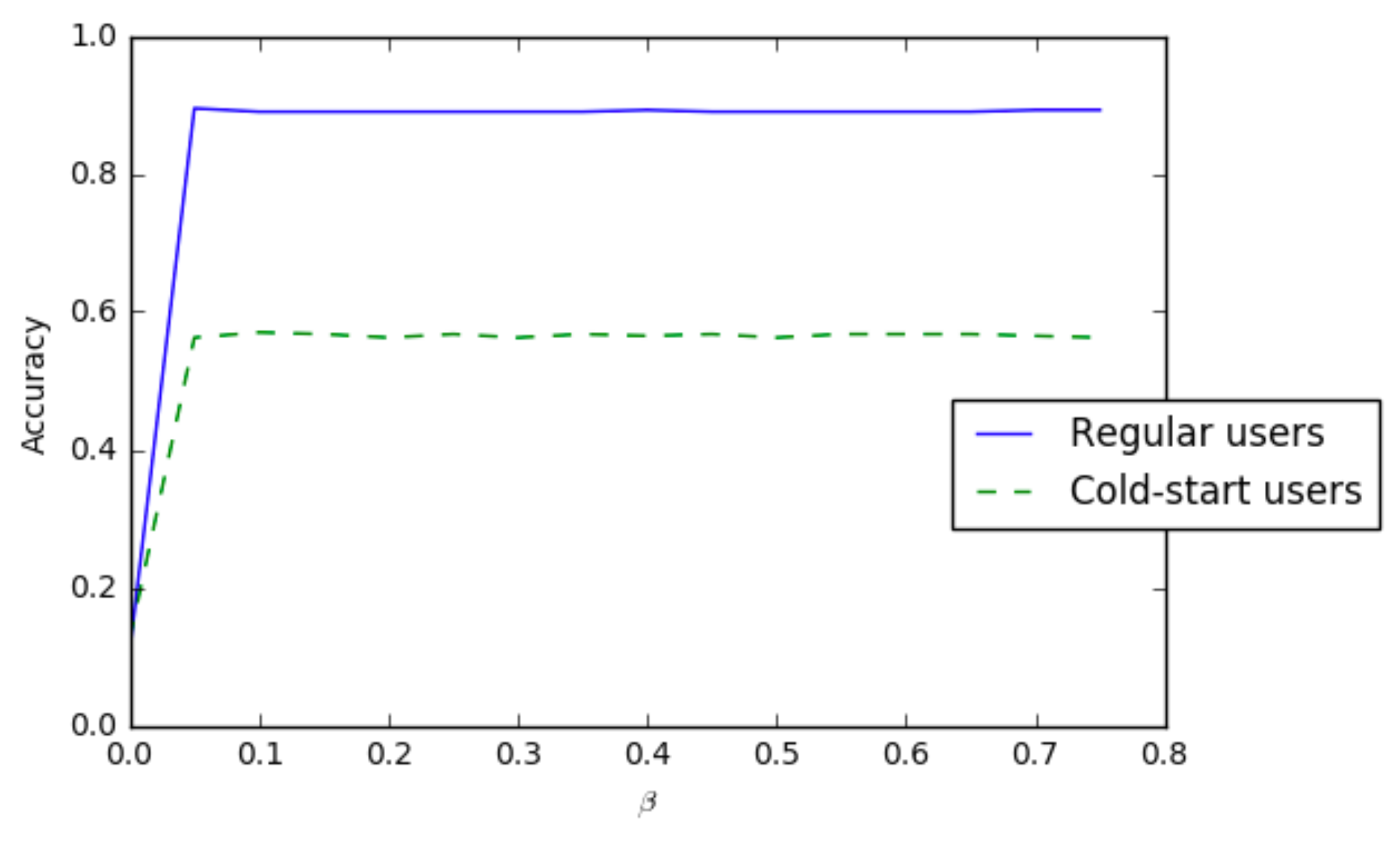}
        \caption{Accuracy as $\beta$ is increased}
    \end{subfigure}%
    ~ 
    \begin{subfigure}[t]{0.5\textwidth}
        \centering
        \includegraphics[height=1.5in,width=2.7in]{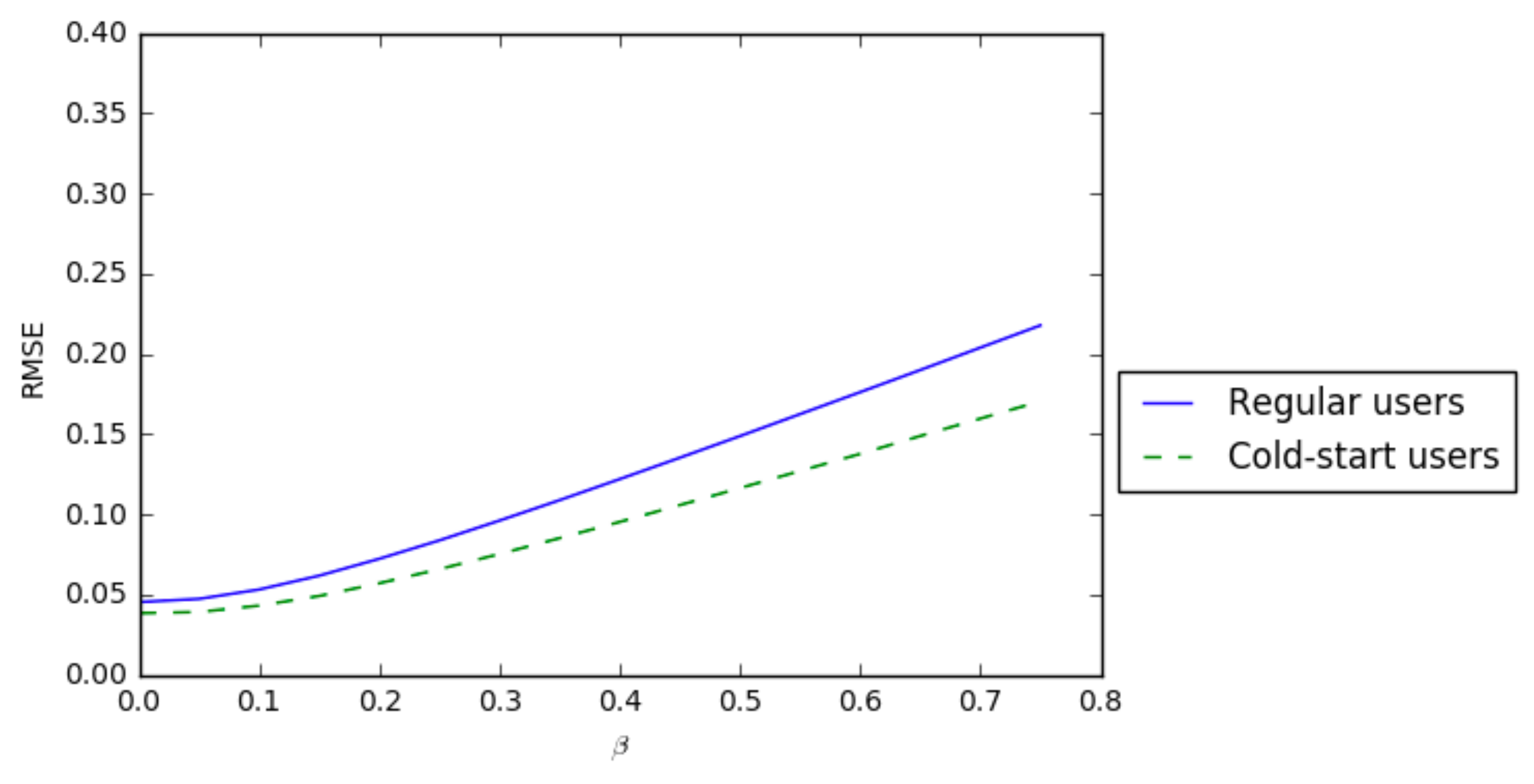}
        \caption{ RMSE as $\beta$ is increased}
    \end{subfigure}
    \caption{Changes in accuracy and RMSE as the impact of the learning profile is increased. \label{fig:learningprofile}}
\end{figure*}

 Figure~\ref{fig:learningprofile} demonstrates how the output of the system changes as we increase $\beta$, which controls the impact of the learning profile. 

 In this figure, $\beta=0$ shows the output of \name where recommendations are made without considering the learning profile (i.e., the system has no knowledge of what topics a question can be classified with). In this setting, the probability of students receiving questions that target their most significant knowledge gap is close to 10\%, which with ten topics is the expected result. As $\beta$ is increased to 0.05, the probability of students receiving questions that target their knowledge gaps is increased without any significant changes in the RMSE. For regular users, the accuracy is 96\% and for cold-start users, the accuracy is close to 57\%. For both sets of users when $\beta > 0.05$, the RMSE starts to increase without significant changes to the accuracy. This suggests that keeping the other parameter settings constant, $\beta=0.05$ produces the best results. This general trend also occurs for other values of $\alpha$ and $L$; however, the accuracy drops as $\alpha$ and $L$ are increased, and the best value for $\beta$ varies in different experiments.

The results of this experiment demonstrate that the value of $\beta$ has a significant impact on the performance of the system. The goal is to set $\beta$, such that users will be exposed to questions targeting their knowledge gaps without making significant sacrifices on the RMSE, which partially represents users' interests.

\begin{figure*}[h]
    \centering
    \begin{subfigure}[t]{0.5\textwidth}
        \centering
        \includegraphics[height=1.5in,width=2.7in]{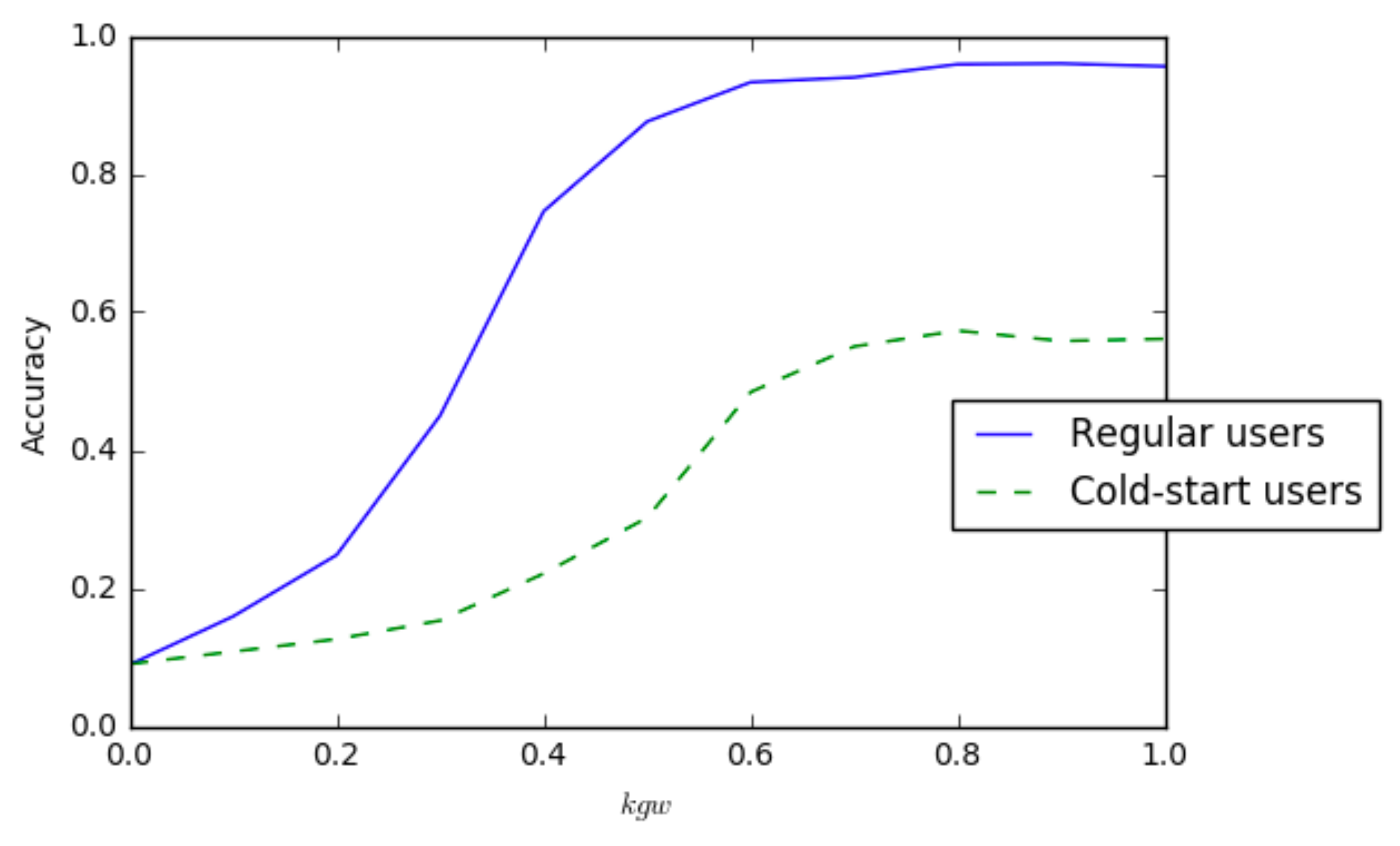}
        \caption{Accuracy as kgw is increased}
    \end{subfigure}%
    ~ 
    \begin{subfigure}[t]{0.5\textwidth}
        \centering
        \includegraphics[height=1.5in,width=2.7in]{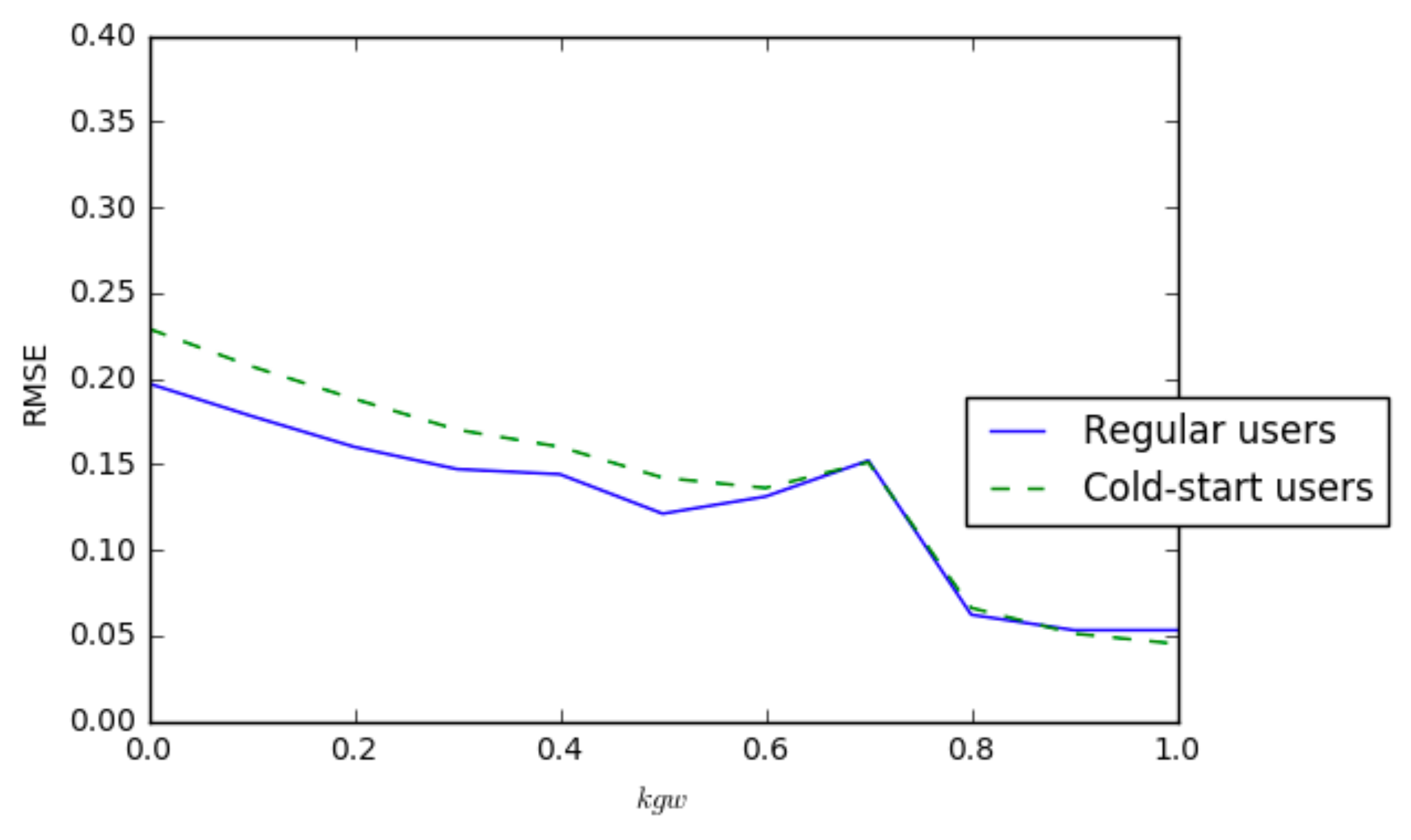}
        \caption{ RMSE as kgw is increased}
    \end{subfigure}
    \caption{Changes in accuracy and RMSE as the impact of knowledge gaps in defining user benefits is increased. \label{fig:kgw}}
\end{figure*}

\paragraph{Impact of Knowledge Gaps in Defining User Benefits ($kgw$):} 
Figure~\ref{fig:kgw} shows how the output of \name changes as we increase $kgw$, which determines how much the system should emphasize knowledge gaps compared to the interests of users.


Increasing $kgw$ improves \name's accuracy in providing users with recommendations that target their most significant gap while reducing the overall RMSE. This is expected since the increase in $kgw$ adds more synergy between users' defined benefits and their knowledge gaps. The sacrifice,  which is not captured by either the accuracy or the RMSE, is that increasing $kgw$ tailors recommendations away from users' interests. With small values of $kgw$, where the emphasis is mostly on the interests, \name is unable to provide recommendations that target user's most significant knowledge gaps. When $kgw$ is set to 0.8, the system has almost 100\% chance of providing regular users with recommendations that target their most significant gap. A similar trend, but with a lower overall accuracy, is observed for cold-start users. Though the system provides the flexibility for the user or a meta-user (e.g., instructors) to set the value of $kgw$, it appears that increasing it beyond a point, which in this experiment is 0.8, may only disregard users' interests without providing any additional benefits in determining gaps.


\paragraph{Summary} The results presented in this section demonstrate that both $kgw$ and $\beta$ can affect the impact of knowledge gaps in the final recommendations made by \name. The main difference is that $kgw$ sets the impact at a question-level in defining how student benefits are determined, which in turn impacts how values in $R$ are stored. In contrast, $\beta$ establishes the impact at a topic-level using the learning profile, ensuring that the system's recommendations are aligned with the user's interests (based on $kgw$). 

\section{Exploration: Using a Historical Data Set}\label{sec:exploration}

The behavior of \name is explored in this section using a historical data set from a first year programming course at The University of British Columbia. The data set is captured in the course by using PeerWise. Two main questions are considered here:

\textbf{Question 1.} How does the accuracy of \name respond as one of its core components, the RecSys, is varied across a collection of standard techniques? The conjecture is that the RecSys is an independent component: the accuracy of \name reflects the accuracy of the RecSys technique used.  In other words \name does not have unintended interactions with the RecSys used, which would allow its replacement in the future as new techniques, with improved accuracy, become available.
 
\textbf{Question 2.} How well do the recommendations made by \name reflect a student's knowledge gaps and interests? The quality of the recommendations are considered using the following refinement:
How do the identified knowledge gaps for users relate to their final examination achievements on topics? The conjecture is that the identified knowledge gaps in \name reflect the actual knowledge gaps of the users as indicated in their final examination achievements. In other words the knowledge gaps identified for users by \name are accurate.  

How do the recommended questions for users relate to their identified knowledge gaps? The conjecture is that the recommended questions in \name would match the identified knowledge gaps of the users. In other words, how accurate is \name when recommending questions?  

Are the recommended questions for users personalized? The conjecture is that if the recommended questions in \name match the identified knowledge gaps of the users and there are a large number of users, then a large number of distinct questions that span the topics would be recommended by \name. This leads to a personalization of  the questions recommended by \name are personalized for individual students.  

Following a description of the historical data set in subsection~\ref{sec:DS}, Question~1 is explored in subsection~\ref{sec:crosscomparison}, and Question~2 is considered in~\ref{sec:report}

\subsection{Data Set Description}
\label{sec:DS}
A historical PeerWise data set has been used in this analysis that was created in a required, introductory course in C programming for engineering students offered at The University of British Columbia in 2014. To encourage participation,  students received grades for their use of the PeerWise environment: (i) They were required to author at least 3 questions and to correctly answer at least 45 questions (worth 1.5\% of final mark), and (ii) a grade was calculated from  the ``Answer Score" (AS) and `` Reputation Score" (RS), which were computed by the PeerWise system, using the following formula: $\frac{min(AS,RS) \times 1.5}{500}$, (worth 1.5\% of final mark). In total 377 students authored 1111 questions, assigned 1700 tags that cover 10 topics, and answered 21432 questions.

Recalling the discussion of Section~\ref{sec:input} it is necessary to scale some of the data stored in the matrices $P$ and $D$ to real values between [0, 1] for this study. The answers from matrix $A$ are binary and do not require scaling. The results reported here are generated using 60\% of data for training the model with different hyper-parameter settings, 20\% of data used for setting the hyper-parameters, and 20\% for assessing the accuracy of the model.

\subsection{Exploring the Behavior of \name Using Alternative RecSys Techniques}\label{sec:crosscomparison}

As Figure~\ref{fig:overview} summarizes, \name uses a RecSys for predicting the benefits users might receive from answering unseen questions. In this study the behavior of \name is explored as the RecSys is varied across a collection of standard techniques. MyMediaLite \cite{Gantner2011}, an open source RecSys library that provides implementations of a collection of standard RecSys algorithms, is used to compare the accuracy of \name the following standard RecSys techniques:
\begin{compactdesc}
\item[User-based Average (U-AVG)] which computes the average ratings across all of the users to approximate how $u$ might rate unseen items. 
 \item[Item-based Average (I-AVG)] which computes the average ratings across all of the items to approximate how $i$ might be rated by users. 
\item[User-based KNN (U-KNN)] computes similarities between users using the Pearson correlation
coefficient to find the $K$ most similar users to a user $u$. The past ratings from the $K$ nearest neighbors are then used to approximate how $u$ might rate unseen items. 
\item[Item-based KNN (I-KNN)] computes similarities between items using the Pearson correlation
coefficient to find the $K$ most similar items to an item $i$. The past ratings that the $K$ nearest neighbors have received are then used to approximate how $i$ might be rated by users. 
\item[Matrix Factorization (MF)] in its simplest form as described in Section~\ref{sec:matrixfactorization}.
\item[Biased Matrix Factorization(BMF)] extends conventional matrix factorization with the addition of a bias parameter for each user and item as described by ~\cite{Koren2008}. 

\end{compactdesc}


Many additional, alternative algorithms are available, which could also have been employed by \name. 
In this work the discussion is limited to solutions available in established, open-source libraries to ensure that developing a prototype system is feasible.

\begin {figure}[h]
\center
\includegraphics[width=8 cm]{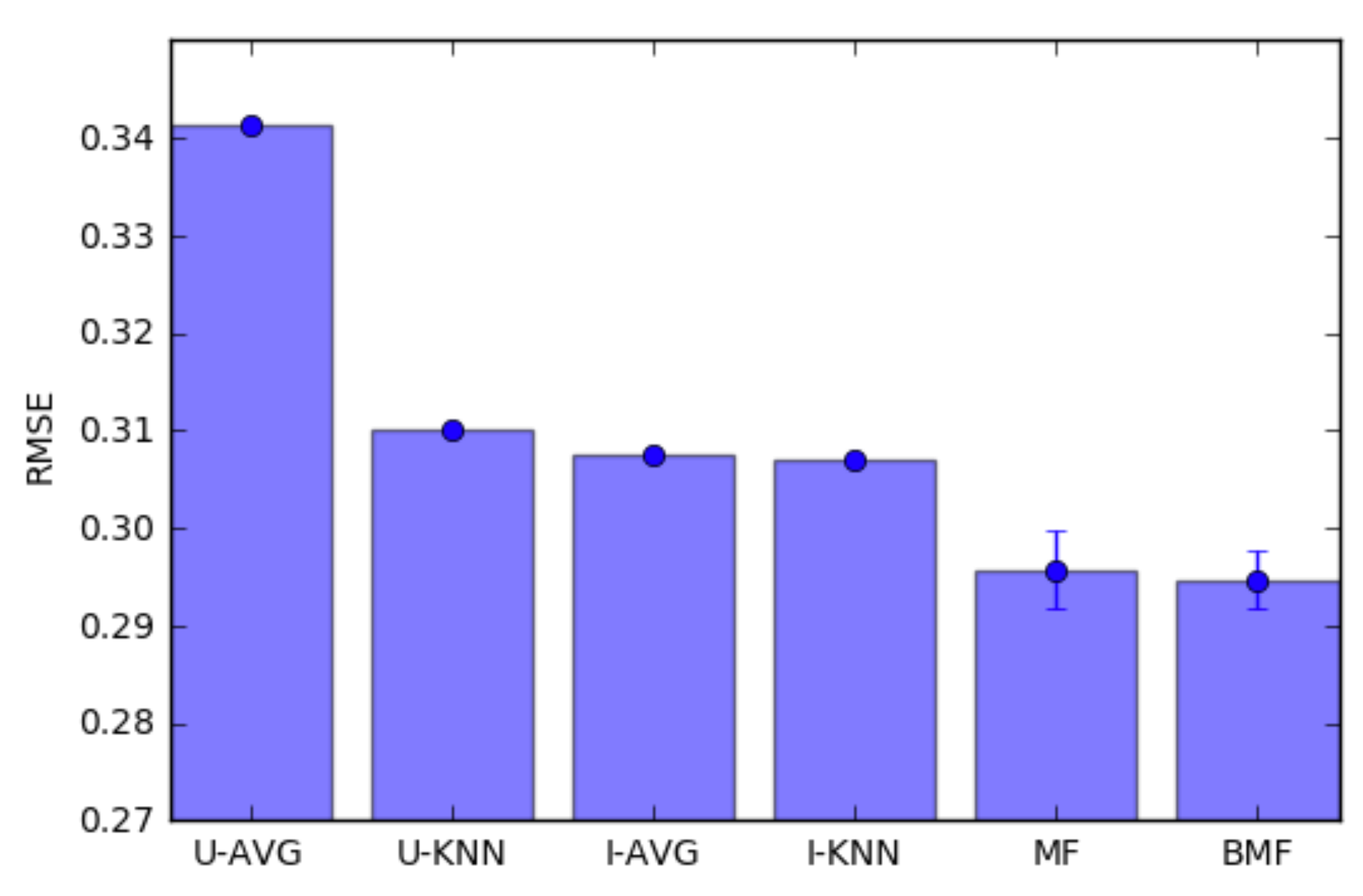}
\caption{Comparison of the predictive accuracy of \name with a number of standard RecSys techniques implemented in MyMediaLite using RMSE \label{fig-results}}
\end {figure}

Figure~\ref{fig-results} visualizes the RMSE results for the cases where the six techniques are used in \name; error bars are calculated using standard deviation. 
$MF$ and $BMF$, which are both based on matrix factorization, outperform the standard user-based or item-based approaches and are within standard error from one another in this data set. Their superior performance can be explained by their ability to implicitly incorporate latent features that may tie to characteristics such as the ``slip" and ``guess" factors \cite{Thai-Nghe2011}.

In response to Question 1, the results indicate the RecSys behaves much like an independent component: the accuracy of \name consistently reflects the accuracy of the RecSys technique used. The RecSys could reasonably be replaced in future studies, for example, as further improvements become available in the community.

\subsection{Exploring the Recommendation Quality by \name}\label{sec:report}


In this study the behavior of \name is explored with respect to the quality of the recommendations produced. The analysis is performed in a single run on the data set, which would correspond to providing recommendations to users in time to prepare for their final examination near the end of the term. Given that the analysis is performed at the end of the term and users receive grades for participation, the data set has only a few cold-start users. Consequently, the study considers the entire cohort without splitting them into regular and cold-start users. 

The study uses \name with the most accurate RecSys technique identified (Section~\ref{sec:crosscomparison}): $BMF$. The following hyper-parameter settings are identified: $\gamma=0.002$, $K=2$, $itr_r= 300$, $kgw = 0.8$, and $\beta= 0.51$. In all cases other than $kgw$, the hyper-parameter values are derived using the validation set; under this setting the RMSE is computed as $0.2947$.  

A summary of the data characteristics, recommendations, and the final examination score achieved is illustrated in Table~\ref{tab:real} (topic name, number of questions, class-level gap, total number of users that received recommendations on each topic, and the class-level exam grade). The topics are listed in the order in which they are covered in class, which may explain why some of the earlier topics receive more questions than others.

\begin{table}[h]
\centering
\begin{tabular}{lrrcc}
\\ \hline 
\multicolumn{1}{c}{\textbf{Topic}} & \multicolumn{1}{l}{\textbf{\# Questions}} & {\textbf{\begin{tabular}[c]{@{}c@{}}Class-level\\  Gap ($\bar{lp}$)\end{tabular}}} & {\textbf{\begin{tabular}[c]{@{}c@{}}\# Topic-level\\ Recommended\end{tabular}}} & \multicolumn{1}{c}{\textbf{\begin{tabular}[c]{@{}l@{}}Class-level \\ Exam Grade\end{tabular}}} \\ \hline
Introduction                       & 175                                       & -0.202                   & 23                                  & 80\%                                                 \\ \hline
Fundamentals                       & 554                                       & 0.012                    & 49                                  & 91\%                                                 \\ \hline
Conditionals                       & 81                                        & -0.023                   & 30                                  & 93\%                                                 \\ \hline
Loops                              & 246                                       & -0.033                   & 34                                  & 68\%                                                 \\ \hline
\textbf{File I/O}                  & 34                                        & 0.160                     & 31                                  & 66\%                                                 \\ \hline
\textbf{Functions}                 & 218                                       & 0.084                    & 41                                  & 55\%                                               \\ \hline
Arrays                             & 276                                       & -0.020                    & 38                                  & 65\%                                                   \\ \hline
DAQ Systems                        & 75                                        & -0.090                    & 6                                   & 80\%                                               \\ \hline
\textbf{Comprehension}             & 15                                        & 0.210                     & 30                                   & 30\%                                                 \\ \hline
Syntax                             & 41                                        & 0.070                     & 9                                   & 77\%       \\ \hline                                         
\end{tabular}
\caption{Information on the name, number of questions, class-level gap, total number of users that received a recommendation on each topic, and the class-level exam grade. \label{tab:real}}
\end{table}

In response to Question 2, the results indicate the quality of the recommendations by \name is promising. With respect to the quality of the identified knowledge gaps by \name, the topic of the three most significant class-level gaps (average gap on topics over the cohort) are programming comprehension, File IO, and functions. These match the topics that receive the lowest average grade on the final exam, indicating the knowledge gaps identified by \name reflect the actual knowledge gaps of the students. 

With respect to the relationship of the recommended questions for users to their identified knowledge gaps, \textbf{89\%} of the users receive recommendations that match their primary (most significant) gap identified from their learning profile vector. In other cases the users receive recommendations on alternative questions that match knowledge gaps in their learning profile vector. This can occur, for example, when the interests of a user are determined to be a superior match for questions that address a secondary or tertiary ranked knowledge gap. 

With respect to the quality of personalized recommendations by \name, the questions recommended at a topic-level broadly span all available topics. In addition, a relatively large number, 99, of distinct questions are presented as the first recommendation to individual students. These results suggest that \name differentiates users and provides personalized recommendations.

\section{Conclusions and Future Work}\label{sec:conclusion}
Explicitly addressing knowledge gaps in RecSysTEL is a challenging, open research topic. A novel RecSysTEL, \name, was introduced as a way of providing accurate, personalized recommendations to students who are using Peer-Learning Environments. The system consists of five modules defining the Input Data, Data Integration, Learning Profile, Recommendation Engine, and Modes of Operation. The Recommendation Engine uses an established collaborative filtering algorithm (matrix factorization), which is enhanced with a learning profile. This enables the recommendation of specific  multiple-choice questions to users that reflect both (i) user interests (the kinds of questions they rank highly), and (ii) their knowledge gaps (the questions they need work on to accomplish learning objectives related to a course). Multiple operational modes allow students to explore new questions, review previously answered questions, or focus on questions from specific topics according to their current learning requirements. 

Experimental validation of \name used both synthetic data sets and a historical data set. The synthetic data sets were used to assess the behavior of \name under diverse circumstances, by varying parameters in both the data generation template and the \name model. This demonstrated that the behavior of \name is consistent with expectations over a range of parameter settings, and therefore that there is reason to believe the solution will be robust in a real world setting. The historical data set was used to explore the accuracy of alternative RecSys techniques, and to demonstrate that \name is likely to both provide recommendations that are both useful and personalized. As \name was able to extract class level knowledge gaps ($\hat{lp}$) which correlated with the final class level exam grade, we have reason to believe that \name is performing appropriately in its profiling of students. Indeed, the result that 89\% of students would have received a recommendation that matches with their most significant knowledge gap suggests that the current parameter settings for \name are appropriate and that it has the potential to improve learning outcomes for students.  Furthermore, the wide range in recommended topics (with 99 distinct questions provided as a recommendation in this test) suggests that \name is not fixating upon a class wide weakness, but providing well personalized advice as to what questions might help a student best. These results are promising, suggesting that students may significantly benefit from exposure to a tool such as \name.

There are a number of limitations in the current work which restrict the generalisability of the results.  The most significant limitation is that the validation is not a controlled experiment that provides compelling evidence of \name's capacity to make recommendations that lead to better learning. To address this limitation an A/B experiment is planned, in which the control group would receive random recommendations and the experimental group would receive recommendations from \name, to determine whether \name's recommendations lead to measurable learning gains. The design of the controlled experiment is envisioned to include the use of a PeerWise environment extended with \name. Discussions are underway with the founder of PeerWise for integrating \name into their platform. 
Furthermore, the historical data set is from one class of students (first year undergraduate engineering course, C programming, Computer Science Department, and so on), which means that current parameter settings are unlikely to generalize across all educational settings. Further investigation is needed to explore the behavior and application of \name in alternative educational domains (e.g., Medicine, Humanities) and settings (e.g., MOOCS). These may, for example, drive the need for very high levels of scalability in terms of the number of users, questions, and course topics.

There are several interesting directions to pursue in future work. The formulation of the updated student-question knowledge gap matrix $\hat{G}_{N\times M}$ 
could be explored as a 
similarity function potentially describing a network among the questions. It would be interesting to compare the results of the different formulations (i.e, learning profile approach vs. a  network approach). It would also be important to refine \name to consider the learning effect using alternative factorization techniques, as students improve their understanding of topics over time. The PeerWise data set provides timestamp information, which may be included in the recommendation model to create more sophisticated models of individual users which evolve in time. Also, the interpretability of the recommendations made by \name is a worthy topic of further investigation in itself. Matrix factorization can result in models that are easier to understand, which suggests that a study could be designed to explore how students make sense of the recommendations, and how this impacts upon their metacognition, and ability to reflect upon and improve their participation in Peer-Learning environments.

In conclusion the results are promising and demonstrate that it is possible to combine MF based learning profiles with a RecSys designed to help students identify knowledge gaps and then work to remove them. The presented system, while designed for the PeerWise environment, is general enough that it could be applied to other environments which store similar information. This means that the system presented here can be used to explore many issues related to student profiling and personalization in a wide variety of question answering scenarios. 

\bibliography{Master-references,kirstybib}  

\begin{thebibliography}{}

\bibitem[\protect\citeauthoryear{Barnes}{Barnes}{2005}]{barnes:qmatrix}
{\sc Barnes, T.} 2005.
\newblock The q-matrix method: Mining student response data for knowledge.
\newblock In {\em American Association for Artificial Intelligence 2005
  Educational Data Mining Workshop}.

\bibitem[\protect\citeauthoryear{Barnes}{Barnes}{2010}]{barnes:novel}
{\sc Barnes, T.} 2010.
\newblock Novel derivation and application of skill matrices: The q-matrix
  method.
\newblock {\em Handbook on educational data mining\/}, 159--172.

\bibitem[\protect\citeauthoryear{Bates, Galloway, McBride, Rebello, Engelhardt,
  and Singh}{Bates et~al\mbox{.}}{2012}]{Bates2012}
{\sc Bates, S.~P.}, {\sc Galloway, R.~K.}, {\sc McBride, K.~L.}, {\sc Rebello,
  N.~S.}, {\sc Engelhardt, P.~V.}, {\sc and} {\sc Singh, C.} 2012.
\newblock Student-generated content: Using peerwise to enhance engagement and
  outcomes in introductory physics courses.
\newblock In {\em AIP Conference Proceedings-American Institute of Physics}.
  Vol. 1413. Citeseer, 123.

\bibitem[\protect\citeauthoryear{Beck and Chang}{Beck and
  Chang}{2007}]{Beck2007}
{\sc Beck, J.~E.} {\sc and} {\sc Chang, K.-m.} 2007.
\newblock Identifiability: A fundamental problem of student modeling.
\newblock In {\em International Conference on User Modeling}. Springer,
  137--146.

\bibitem[\protect\citeauthoryear{Beswick, Rothblum, and Mann}{Beswick
  et~al\mbox{.}}{1988}]{beswick.rothblum.ea:psychological}
{\sc Beswick, G.}, {\sc Rothblum, E.~D.}, {\sc and} {\sc Mann, L.} 1988.
\newblock Psychological antecedents of student procrastination.
\newblock {\em Australian psychologist\/}~{\em 23,\/}~2, 207--217.

\bibitem[\protect\citeauthoryear{BETTS}{BETTS}{2013}]{BETTS2013}
{\sc BETTS, B.} 2013.
\newblock Towards a method of improving participation in online collaborative
  learning: Curatr.
\newblock {\em Teaching and Learning Online: New Models of Learning for a
  Connected World\/}~{\em 2}, 260.

\bibitem[\protect\citeauthoryear{Biggs}{Biggs}{1999}]{biggs:what}
{\sc Biggs, J.} 1999.
\newblock What the student does: teaching for enhanced learning.
\newblock {\em Higher education research \& development\/}~{\em 18,\/}~1,
  57--75.

\bibitem[\protect\citeauthoryear{Boud, Cohen, and Sampson}{Boud
  et~al\mbox{.}}{2014}]{Boud2014}
{\sc Boud, D.}, {\sc Cohen, R.}, {\sc and} {\sc Sampson, J.} 2014.
\newblock {\em Peer learning in higher education: Learning from and with each
  other}.
\newblock Routledge.

\bibitem[\protect\citeauthoryear{Cazella, Reategui, and Behar}{Cazella
  et~al\mbox{.}}{2010}]{Cazella2010}
{\sc Cazella, S.}, {\sc Reategui, E.}, {\sc and} {\sc Behar, P.} 2010.
\newblock Recommendation of learning objects applying collaborative filtering
  and competencies.
\newblock In {\em Key Competencies in the Knowledge Society}. Springer, 35--43.

\bibitem[\protect\citeauthoryear{Cechinel, Sicilia, S{\'a}Nchez-Alonso, and
  Garc{\'\i}A-Barriocanal}{Cechinel et~al\mbox{.}}{2013}]{Cechinel2013}
{\sc Cechinel, C.}, {\sc Sicilia, M.-{\'A}.}, {\sc S{\'a}Nchez-Alonso, S.},
  {\sc and} {\sc Garc{\'\i}A-Barriocanal, E.} 2013.
\newblock Evaluating collaborative filtering recommendations inside large
  learning object repositories.
\newblock {\em Information Processing \& Management\/}~{\em 49,\/}~1, 34--50.

\bibitem[\protect\citeauthoryear{Chen}{Chen}{2008}]{chen2008}
{\sc Chen, C.-M.} 2008.
\newblock Intelligent web-based learning system with personalized learning path
  guidance.
\newblock {\em Computers \& Education\/}~{\em 51,\/}~2, 787--814.

\bibitem[\protect\citeauthoryear{Chin and Brown}{Chin and
  Brown}{2002}]{Chin2002}
{\sc Chin, C.} {\sc and} {\sc Brown, D.~E.} 2002.
\newblock Student-generated questions: A meaningful aspect of learning in
  science.
\newblock {\em International Journal of Science Education\/}~{\em 24,\/}~5,
  521--549.

\bibitem[\protect\citeauthoryear{Coetzee, Lim, Fox, Hartmann, and
  Hearst}{Coetzee et~al\mbox{.}}{2015}]{Coetzee2015}
{\sc Coetzee, D.}, {\sc Lim, S.}, {\sc Fox, A.}, {\sc Hartmann, B.}, {\sc and}
  {\sc Hearst, M.~A.} 2015.
\newblock Structuring interactions for large-scale synchronous peer learning.
\newblock In {\em Proceedings of the 18th ACM Conference on Computer Supported
  Cooperative Work \& Social Computing}. ACM, 1139--1152.

\bibitem[\protect\citeauthoryear{d~Baker, Corbett, and Aleven}{d~Baker
  et~al\mbox{.}}{2008}]{Baker2008}
{\sc d~Baker, R.~S.}, {\sc Corbett, A.~T.}, {\sc and} {\sc Aleven, V.} 2008.
\newblock More accurate student modeling through contextual estimation of slip
  and guess probabilities in bayesian knowledge tracing.
\newblock In {\em International Conference on Intelligent Tutoring Systems}.
  Springer, 406--415.

\bibitem[\protect\citeauthoryear{Denny, Hamer, Luxton-Reilly, and
  Purchase}{Denny et~al\mbox{.}}{2008}]{Denny2008}
{\sc Denny, P.}, {\sc Hamer, J.}, {\sc Luxton-Reilly, A.}, {\sc and} {\sc
  Purchase, H.} 2008.
\newblock Peerwise: students sharing their multiple choice questions.
\newblock In {\em Proceedings of the fourth international workshop on computing
  education research}. ACM, 51--58.

\bibitem[\protect\citeauthoryear{Desmarais}{Desmarais}{2012}]{desmarais:mapping}
{\sc Desmarais, M.~C.} 2012.
\newblock Mapping question items to skills with non-negative matrix
  factorization.
\newblock {\em ACM SIGKDD Explorations Newsletter\/}~{\em 13,\/}~2, 30--36.

\bibitem[\protect\citeauthoryear{Desmarais, Beheshti, and Naceur}{Desmarais
  et~al\mbox{.}}{2012}]{desmarais.beheshti.ea:item}
{\sc Desmarais, M.~C.}, {\sc Beheshti, B.}, {\sc and} {\sc Naceur, R.} 2012.
\newblock {\em Item to Skills Mapping: Deriving a Conjunctive Q-matrix from
  Data}.
\newblock Springer Berlin Heidelberg, Berlin, Heidelberg, 454--463.

\bibitem[\protect\citeauthoryear{Desmarais and Naceur}{Desmarais and
  Naceur}{2013}]{desmarais.naceur:matrix}
{\sc Desmarais, M.~C.} {\sc and} {\sc Naceur, R.} 2013.
\newblock {\em A Matrix Factorization Method for Mapping Items to Skills and
  for Enhancing Expert-Based Q-Matrices}.
\newblock Springer Berlin Heidelberg, Berlin, Heidelberg, 441--450.

\bibitem[\protect\citeauthoryear{Desmarais and Pelczer}{Desmarais and
  Pelczer}{2010}]{Desmarais2010}
{\sc Desmarais, M.~C.} {\sc and} {\sc Pelczer, I.} 2010.
\newblock On the faithfulness of simulated student performance data.
\newblock In {\em Educational Data Mining 2010}.

\bibitem[\protect\citeauthoryear{Drachsler, Verbert, Santos, and
  Manouselis}{Drachsler et~al\mbox{.}}{2015}]{Drachsler2015}
{\sc Drachsler, H.}, {\sc Verbert, K.}, {\sc Santos, O.~C.}, {\sc and} {\sc
  Manouselis, N.} 2015.
\newblock Panorama of recommender systems to support learning.
\newblock In {\em Recommender systems handbook}. Springer, 421--451.

\bibitem[\protect\citeauthoryear{Drasgow and Hulin}{Drasgow and
  Hulin}{1990}]{Drasgow1990}
{\sc Drasgow, F.} {\sc and} {\sc Hulin, C.~L.} 1990.
\newblock Item response theory.

\bibitem[\protect\citeauthoryear{Erdt, Fern{\'a}ndez, and Rensing}{Erdt
  et~al\mbox{.}}{2015}]{Erdt2015}
{\sc Erdt, M.}, {\sc Fern{\'a}ndez, A.}, {\sc and} {\sc Rensing, C.} 2015.
\newblock Evaluating recommender systems for technology enhanced learning: A
  quantitative survey.
\newblock {\em IEEE Transactions on Learning Technologies\/}~{\em 8,\/}~4,
  326--344.

\bibitem[\protect\citeauthoryear{Fazeli, Loni, Drachsler, and Sloep}{Fazeli
  et~al\mbox{.}}{2014}]{Fazeli2014}
{\sc Fazeli, S.}, {\sc Loni, B.}, {\sc Drachsler, H.}, {\sc and} {\sc Sloep,
  P.} 2014.
\newblock Which recommender system can best fit social learning platforms?
\newblock In {\em European Conference on Technology Enhanced Learning}.
  Springer, 84--97.

\bibitem[\protect\citeauthoryear{Gantner, Rendle, Freudenthaler, and
  Schmidt-Thieme}{Gantner et~al\mbox{.}}{2011}]{Gantner2011}
{\sc Gantner, Z.}, {\sc Rendle, S.}, {\sc Freudenthaler, C.}, {\sc and} {\sc
  Schmidt-Thieme, L.} 2011.
\newblock Mymedialite: a free recommender system library.
\newblock In {\em Proceedings of the fifth ACM conference on Recommender
  systems}. ACM, 305--308.

\bibitem[\protect\citeauthoryear{Gomez-Albarran and
  Jimenez-Diaz}{Gomez-Albarran and Jimenez-Diaz}{2009}]{Gomez-Albarran2009}
{\sc Gomez-Albarran, M.} {\sc and} {\sc Jimenez-Diaz, G.} 2009.
\newblock Recommendation and students' authoring in repositories of learning
  objects: A case-based reasoning approach.
\newblock {\em International Journal of Emerging Technologies in Learning\/}.

\bibitem[\protect\citeauthoryear{Hardy, Bates, Casey, Galloway, Galloway, Kay,
  Kirsop, and McQueen}{Hardy et~al\mbox{.}}{2014}]{Hardy2014}
{\sc Hardy, J.}, {\sc Bates, S.~P.}, {\sc Casey, M.~M.}, {\sc Galloway, K.~W.},
  {\sc Galloway, R.~K.}, {\sc Kay, A.~E.}, {\sc Kirsop, P.}, {\sc and} {\sc
  McQueen, H.~A.} 2014.
\newblock Student-generated content: Enhancing learning through sharing
  multiple-choice questions.
\newblock {\em International Journal of Science Education\/}~{\em 36,\/}~13,
  2180--2194.

\bibitem[\protect\citeauthoryear{Imran, Belghis-Zadeh, Chang, Graf,
  et~al\mbox{.}}{Imran et~al\mbox{.}}{2016}]{Imran2016}
{\sc Imran, H.}, {\sc Belghis-Zadeh, M.}, {\sc Chang, T.-W.}, {\sc Graf, S.},
  {\sc et~al\mbox{.}} 2016.
\newblock Plors: a personalized learning object recommender system.
\newblock {\em Vietnam Journal of Computer Science\/}~{\em 3,\/}~1, 3--13.

\bibitem[\protect\citeauthoryear{Kopeinik, Lex, Seitlinger, Albert, and
  Ley}{Kopeinik et~al\mbox{.}}{2017}]{Kopeinik2017}
{\sc Kopeinik, S.}, {\sc Lex, E.}, {\sc Seitlinger, P.}, {\sc Albert, D.}, {\sc
  and} {\sc Ley, T.} 2017.
\newblock Supporting collaborative learning with tag recommendations: a
  real-world study in an inquiry-based classroom project.
\newblock In {\em Proceedings of the Seventh International Learning Analytics
  \& Knowledge Conference}. ACM, 409--418.

\bibitem[\protect\citeauthoryear{Koren}{Koren}{2008}]{Koren2008}
{\sc Koren, Y.} 2008.
\newblock Factorization meets the neighborhood: a multifaceted collaborative
  filtering model.
\newblock In {\em Proceedings of the 14th ACM SIGKDD international conference
  on Knowledge discovery and data mining}. ACM, 426--434.

\bibitem[\protect\citeauthoryear{Lemire, Boley, McGrath, and Ball}{Lemire
  et~al\mbox{.}}{2005}]{Lemire2005}
{\sc Lemire, D.}, {\sc Boley, H.}, {\sc McGrath, S.}, {\sc and} {\sc Ball, M.}
  2005.
\newblock Collaborative filtering and inference rules for context-aware
  learning object recommendation.
\newblock {\em Interactive Technology and Smart Education\/}~{\em 2,\/}~3,
  179--188.

\bibitem[\protect\citeauthoryear{Lumezanu, Levin, and Spring}{Lumezanu
  et~al\mbox{.}}{2007}]{Lumezanu2007}
{\sc Lumezanu, C.}, {\sc Levin, D.}, {\sc and} {\sc Spring, N.} 2007.
\newblock Peerwise discovery and negotiation of faster paths.
\newblock In {\em HotNets}. Citeseer.

\bibitem[\protect\citeauthoryear{Mangina and Kilbride}{Mangina and
  Kilbride}{2008}]{Mangina2008}
{\sc Mangina, E.} {\sc and} {\sc Kilbride, J.} 2008.
\newblock Evaluation of keyphrase extraction algorithm and tiling process for a
  document/resource recommender within e-learning environments.
\newblock {\em Computers \& Education\/}~{\em 50,\/}~3, 807--820.

\bibitem[\protect\citeauthoryear{Manouselis, Drachsler, Vuorikari, Hummel, and
  Koper}{Manouselis et~al\mbox{.}}{2011}]{Manouselis2011}
{\sc Manouselis, N.}, {\sc Drachsler, H.}, {\sc Vuorikari, R.}, {\sc Hummel,
  H.}, {\sc and} {\sc Koper, R.} 2011.
\newblock Recommender systems in technology enhanced learning.
\newblock In {\em Recommender systems handbook}. Springer, 387--415.

\bibitem[\protect\citeauthoryear{Pardos and Heffernan}{Pardos and
  Heffernan}{2010}]{Pardos2010}
{\sc Pardos, Z.~A.} {\sc and} {\sc Heffernan, N.~T.} 2010.
\newblock Using hmms and bagged decision trees to leverage rich features of
  user and skill from an intelligent tutoring system dataset.
\newblock {\em Journal of Machine Learning Research W \& CP\/}.

\bibitem[\protect\citeauthoryear{Purchase, Hamer, Denny, and
  Luxton-Reilly}{Purchase et~al\mbox{.}}{2010}]{Purchase2010}
{\sc Purchase, H.}, {\sc Hamer, J.}, {\sc Denny, P.}, {\sc and} {\sc
  Luxton-Reilly, A.} 2010.
\newblock The quality of a peerwise mcq repository.
\newblock In {\em Proceedings of the Twelfth Australasian Conference on
  Computing Education-Volume 103}. Australian Computer Society, Inc., 137--146.

\bibitem[\protect\citeauthoryear{Ricci, Rokach, and Shapira}{Ricci
  et~al\mbox{.}}{2011}]{Ricci2011}
{\sc Ricci, F.}, {\sc Rokach, L.}, {\sc and} {\sc Shapira, B.} 2011.
\newblock {\em Introduction to Recommender Systems Handbook}.
\newblock Springer US, Boston, MA, 1--35.

\bibitem[\protect\citeauthoryear{Rosenshine, Meister, and Chapman}{Rosenshine
  et~al\mbox{.}}{1996}]{Rosenshine1996}
{\sc Rosenshine, B.}, {\sc Meister, C.}, {\sc and} {\sc Chapman, S.} 1996.
\newblock Teaching students to generate questions: A review of the intervention
  studies.
\newblock {\em Review of educational research\/}~{\em 66,\/}~2, 181--221.

\bibitem[\protect\citeauthoryear{Salehi}{Salehi}{2013}]{Salehi2013}
{\sc Salehi, M.} 2013.
\newblock Application of implicit and explicit attribute based collaborative
  filtering and bide for learning resource recommendation.
\newblock {\em Data \& Knowledge Engineering\/}~{\em 87}, 130--145.

\bibitem[\protect\citeauthoryear{Thai-Nghe, Drumond, Horv{\'a}th,
  Krohn-Grimberghe, Nanopoulos, and Schmidt-Thieme}{Thai-Nghe
  et~al\mbox{.}}{2011}]{Thai-Nghe2011}
{\sc Thai-Nghe, N.}, {\sc Drumond, L.}, {\sc Horv{\'a}th, T.}, {\sc
  Krohn-Grimberghe, A.}, {\sc Nanopoulos, A.}, {\sc and} {\sc Schmidt-Thieme,
  L.} 2011.
\newblock Factorization techniques for predicting student performance.
\newblock {\em Educational Recommender Systems and Technologies: Practices and
  Challenges\/}, 129--153.

\bibitem[\protect\citeauthoryear{Thai-Nghe, Horv{\'a}th, and
  Schmidt-Thieme}{Thai-Nghe
  et~al\mbox{.}}{2010}]{thai-nghe.horvath.ea:factorization}
{\sc Thai-Nghe, N.}, {\sc Horv{\'a}th, T.}, {\sc and} {\sc Schmidt-Thieme, L.}
  2010.
\newblock Factorization models for forecasting student performance.
\newblock In {\em Educational Data Mining 2011}.

\bibitem[\protect\citeauthoryear{Verbert, Drachsler, Manouselis, Wolpers,
  Vuorikari, and Duval}{Verbert et~al\mbox{.}}{2011}]{Verbert2011}
{\sc Verbert, K.}, {\sc Drachsler, H.}, {\sc Manouselis, N.}, {\sc Wolpers,
  M.}, {\sc Vuorikari, R.}, {\sc and} {\sc Duval, E.} 2011.
\newblock Dataset-driven research for improving recommender systems for
  learning.
\newblock In {\em Proceedings of the 1st International Conference on Learning
  Analytics and Knowledge}. ACM, 44--53.

\bibitem[\protect\citeauthoryear{Winters}{Winters}{2006}]{winters:educational}
{\sc Winters, T.~e.} 2006.
\newblock Educational data mining: collection and analysis of score matrices
  for outcomes-based assessment.
\newblock Ph.D. thesis, Citeseer.

\bibitem[\protect\citeauthoryear{Wright, Thornton, and Leyton-Brown}{Wright
  et~al\mbox{.}}{2015}]{Wright2015}
{\sc Wright, J.~R.}, {\sc Thornton, C.}, {\sc and} {\sc Leyton-Brown, K.} 2015.
\newblock Mechanical ta: Partially automated high-stakes peer grading.
\newblock In {\em Proceedings of the 46th ACM Technical Symposium on Computer
  Science Education}. ACM, 96--101.

\end{thebibliography}
\bibliographystyle{acmtrans}

\end{document}